\documentclass[11pt,aip]{article}
\usepackage{extsizes}
\usepackage[super,sort&compress,comma]{natbib} 
\usepackage[version=3]{mhchem}
\usepackage[left=1.5cm, right=1.5cm, top=1.785cm, bottom=2.0cm]{geometry}
\usepackage{balance}
\usepackage{siunitx}
\usepackage{mathptmx}
\usepackage{sectsty}
\usepackage{graphicx} 
\usepackage{lastpage}
\usepackage[format=plain,justification=justified,singlelinecheck=false,font={stretch=1.125,small,sf},labelfont=bf,labelsep=space]{caption}
\usepackage{float}
\usepackage{fancyhdr}
\usepackage{fnpos}
\usepackage[english]{babel}

\usepackage{array}
\usepackage{droidsans}
\usepackage{charter}
\usepackage[T1]{fontenc}
\usepackage[usenames,dvipsnames]{xcolor}
\usepackage{setspace}
\usepackage[compact]{titlesec}
\usepackage{hyperref}

\usepackage{epstopdf}
\definecolor{cream}{RGB}{222,217,201}

\begin{document}

\thispagestyle{plain}


 \noindent\LARGE{\textbf {\\ Statistics makes a difference: Machine learning adsorption dynamics of functionalized cyclooctine on Si(001) at DFT accuracy\\}
} 

 \noindent\large{Hendrik Weiske,\textit{$^{a}$} Rhyan Barrett,\textit{$^{a}$} Ralf Tonner-Zech,\textit{$^{a}$} Patrick Melix,\textit{$^{a}$} and Julia Westermayr$^{\ast}$\textit{$^{a,b}$ \\}} \\
\textit{$^{a}$~Wilhelm Ostwald Institute, Leipzig University, Leipzig, Germany\\ $^{b}$~Center for Scalable Data Analytics and Artificial Intelligence Dresden/Leipzig, Leipzig, Germany\\ E-mail: julia.westermayr@uni-leipzig.de\\~\\ }
\noindent\normalsize{

The interpretation of experiments on reactive semiconductor surfaces requires statistically significant sampling of molecular dynamics, but conventional ab initio methods are limited due to prohibitive computational costs. Machine-learning interatomic potentials provide a promising solution, bridging the gap between the chemical accuracy of short ab initio molecular dynamics (AIMD) and the extensive sampling required to simulate experiment. Using ethinyl-functionalized cyclooctyne adsorption on Si(001) as a model system, we demonstrate that conventional AIMD undersamples the configurational space, resulting in discrepancies with scanning tunnelling microscopy and X-ray photoelectron spectroscopy data. To resolve these inconsistencies, we employ pre-trained equivariant message-passing neural networks, fine-tuned on only a few thousand AIMD snapshots, and integrate them into a “molecular-gun” workflow. This approach generates 10\,000 independent trajectories more than 1\,000 times faster than AIMD. These simulations recover rare intermediates, clarify the competition between adsorption motifs, and reproduce the experimentally dominant on‑top [2+2] cycloaddition geometry. Our results show that fine-tuning of pre-trained foundational models enables statistically converged, chemically accurate simulations of bond‑forming and bond‑breaking events on complex surfaces, providing a scalable route to reconcile atomistic theory with experimental ensemble measurements in semiconductor functionalization.
}\\






\renewcommand{\headrulewidth}{0pt}


\makeFNbottom
\makeatletter
\renewcommand\LARGE{\@setfontsize\LARGE{15pt}{17}}
\renewcommand\Large{\@setfontsize\Large{12pt}{14}}
\renewcommand\large{\@setfontsize\large{10pt}{12}}
\renewcommand\footnotesize{\@setfontsize\footnotesize{7pt}{10}}
\makeatother
\section{Introduction}

\begin{figure}
    \centering
    \includegraphics[width=0.5\columnwidth]{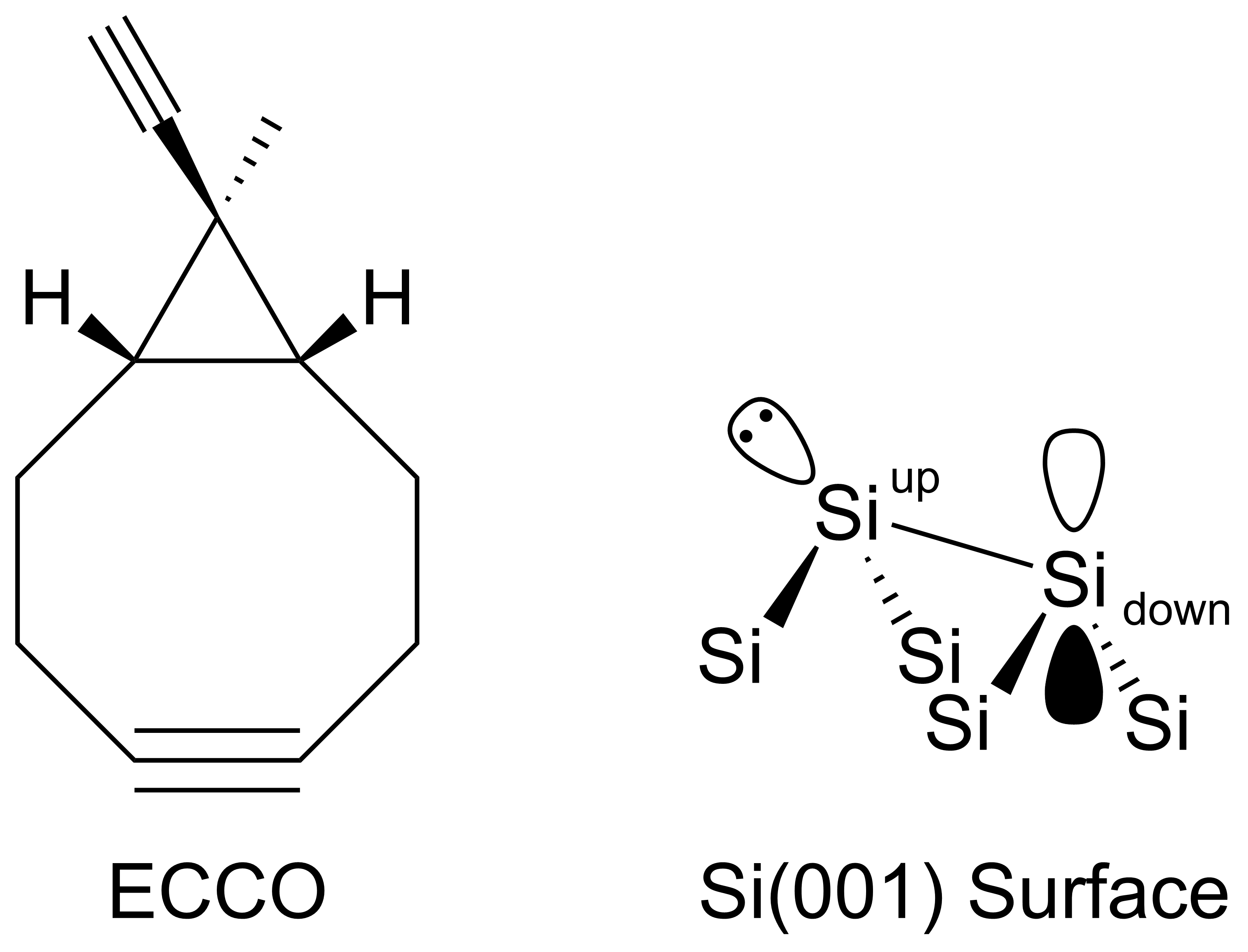}
    \caption{Left: Lewis structure of 9-Ethinyl-9Methylbicyclo[6.1.0]non-4-in (ECCO). Right: reconstructed Si(001) surface, where Si-dimers are formed on the surface consisting of \ce{Si_{up}} and \ce{Si_{down}} atoms. The dimer can be described as a \ce{Si_{up}} atom with a lone pair and a partial negative charge, whereas the \ce{Si_{down}} can be described as carrying an empty p-orbital.}
    \label{fig:lewis_struct}
\end{figure}

The functionalization of semiconductor surfaces, particularly silicon, offers a versatile means to tailor electronic, chemical, and mechanical properties.\cite{Wolkow1999, Teplyakov2013a, Bent2002, Kachian2010} Cyclooctynes, widely used in strain-promoted click chemistry,\cite{Kolb2001,Munster2016,Reutzel2016,Glaser2021, Glaser2021a, Nalaoh2025, Durr2025} serve as selective and reactive agents for Si(001) functionalization, enabling mild, covalent attachment while minimizing side reactions.\cite{Mette2013, Reutzel2016, Pecher2017c, Pecher2017a, Pecher2018b, Pecher2018,Glaser2020, Glaser2021, Glaser2021a, Glaser2024, Langer2019} Surface‐sensitive experimental techniques such as scanning tunneling microscopy (STM) or X‐ray photoelectron spectroscopy (XPS) provide rich detail on adsorption structures, coverage, and side reactions,\cite{Langer2019} yet they lack the temporal and atomistic resolution needed to observe transient intermediates and adsorption pathways required to resolve reaction kinetics.
Moreover, ensemble‐averaged spectroscopies yield information on overall surface composition and functional group identity, but fail to resolve site‐specific energetics or orientation distributions. As a result, critical details, including the relative barriers for adsorption on the two non-equivalent dangling bonds of the Si(001) row (see Figure~\ref{fig:lewis_struct}), the influence of subsurface strain on cyclooctyne ring opening, and the lifetimes of metastable precursors, remain experimentally inaccessible.

Computational approaches can complement experiment, but rely on computationally costly quantum-chemical calculations. As a consequence, studies are often left with static analyses using density functional theory (DFT), which is usually the workhorse of such simulations. \cite{Maurer2016, Maurer2019} However, for capturing reaction kinetics and dynamical processes, molecular dynamics (MD) simulations are needed. Classical force fields, which offer a computationally viable solution, lack the ability to describe covalent bond formation and breaking. For some specific systems, reactive Force Fields (ReaxFF\cite{VanDuin2001, Senftle2016}) have been used in surface chemical studies. However, for statistically relevant sampling, these methods are also too demanding.\cite{Hu2017b, Zhu2020, Wen2017, Nayir2019}
\textit{Ab-initio} molecular dynamics (AIMD), in principle, offers both reactivity and accuracy, yet its computational cost severely limits accessible timescales and statistical sampling.\cite{Iftimie2005, Radeke1997,Grotendorst2000a,Gross2023,Pecher2017c}
Recent work on ethinyl-functionalized cyclooctyne (ECCO, see Figure~\ref{fig:lewis_struct}) adsorption at Si(001) surfaces revealed a bottleneck: AIMD trajectories, even tens of picoseconds in length, can miss key binding modes observed experimentally, leading to discrepancies in predicted versus measured dominant adsorption geometries.\cite{Pieck2021,Langer2019} Whether such mismatches stem from methodological limitations or from simple undersampling remains an open and critical question.

To address this question, we leverage machine learning (ML) to vastly accelerate surface MD simulations without compromising \textit{ab-initio} accuracy. Specifically, we fine-tune the foundational equivariant, message‑passing atomic cluster expansion (MACE) model\cite{Batatia2022, Batatia2024}, MACE-MP-0, using our previous AIMD data for ECCO/Si(001),\cite{Pieck2021} deploying a "molecular gun" strategy that generates thousands of statistically independent trajectories in a black-box fashion.

Therefore, simulations at near-DFT accuracy become accessible, while being multiple orders of magnitude faster.\cite{Stark2024} While the MACE-MP-0 architecture is widely adopted across molecular and materials applications,\cite{Giese2025,Shiota2024,Schaaf2024,Sauer2025,Vanita2025,Singh2024,AbdulNasir2025,Kabylda2025,Demeyere2025,Loew2025,Cheng2025,Shen2024,Schafer2025,Novelli2025,Hormann2025,Schwalbe-Koda2025,Gupta2025,Fako2025,Tian2025,Pitfield2025,Christiansen2025,Soyemi2025,Cvitkovich2024} its suitability for surface fine-tuning with limited data,\cite{Deng2025,Rensmeyer2025,Bertani2025,Hanseroth2025,Kaur2025,Elena2025,Lim2025,Steffen2025,Focassio2025} as demonstrated here, presents a practical solution to statistical convergence issues in surface chemistry.\cite{Allam2024,Boulangeot2024,Du2025}


By fine-tuning a pre-trained MACE-MP-0 model\cite{Batatia2024} with targeted ECCO/Si(001) AIMD snapshots,\cite{Pieck2021} we remove the sampling bottleneck, enabling large-scale, chemically accurate simulations at affordable computational cost. Our machine learning molecular gun allows for detailed analysis of binding-site populations, desorption barriers, and ring-opening dynamics, placing the atomistic mechanism of ECCO adsorption in direct, quantitative correspondence with STM and XPS data.\cite{Langer2019}
\section{Computational details}
To conduct ML-accelerated AIMD, we use the foundational MACE model for materials, MACE-MP-0,\cite{Batatia2024} and fine-tune it on data obtained by some of us in a recent study.\cite{Pieck2021} We therefore only briefly summarise the quantum-chemical reference simulations and the model architecture, referring to the cited publications for full details.

\subsection{AIMD Reference Simulations}
The \textit{ab-initio} data for training were taken from previous work\cite{Pieck2021, Pieck2021a} using DFT-based MD. Trajectories were generated using VASP 5.4.4,\cite{Kresse1993, Kresse1994, Kresse1996, Kresse1996a, Kresse1999} using the exchange-correlation functional by Perdew, Burke, and Ernzerhof (PBE)\cite{Perdew1996, Perdew1997} with the DFT-D3(BJ) dispersion correction scheme.\cite{Grimme2010, Grimme2011} The simulations were designed to model ultra-high vacuum (UHV) deposition experiments, in which evaporated molecules impinge on a surface with finite kinetic energy. We therefore refer to this approach as the “molecular gun”. In this protocol, the Si(001) slab and the ECCO molecule were first equilibrated separately for 40\,ps at 300\,K in the NVT ensemble. From these trajectories, a configuration (coordinates and velocities) was extracted every 1\,ps to sample thermally excited states of both subsystems. Ten such configurations served as initial states for subsequent MD runs, in which the molecule was accelerated towards the surface by adding a random downward velocity component, mimicking the conditions of UHV deposition. We use this same strategy in our work to generate additional initial conditions and improve statistical sampling. 

In the reference data, dynamics were simulated in the NVT ensemble using a Nos\'e–Hoover thermostat\cite{Nose1984,Hoover1985,Nose1991} at 300K with a Nos\'e mass of 1.8.\cite{Pieck2021}

The complete AIMD dataset comprises approximately 327k frames: formed from $\sim$199k frames of the ECCO molecule and $\sim$128k frames of molecular gun runs (ECCO+Si(001)).\cite{Pieck2021a}
All frames in which unphysical C–H bond fission occurred were removed. This was the case in two of the 10 AIMD trajectories.\cite{Pieck2021}
After randomising the remaining frames, every 25\textsuperscript{th} configuration was selected to form the production machine learning dataset (coordinates, velocities, and energies), resulting in $\sim$13\,000 data points of ECCO on Si(001). As shown previously, only a fraction of the entire trajectories is enough to achieve good training results.\cite{Tiefenbacher2025, Smith2018}  
\subsection{Machine Learning MD}
All MD simulations in this work were performed using the MD driver implemented in the atomic simulation environment (ASE).\cite{Larsen2017} As initial configurations, we used the ten starting structures from the reference AIMD simulations,\cite{Pieck2021} providing pre-equilibrated systems (see also subsection 2.1). For each run, the position of the ECCO molecule in the $x$–$y$ plane (parallel to the surface) was randomised. The distance between the surface atom plane and the ECCO centre of mass was fixed at 20\,\AA, corresponding to an approximate shortest atom–surface separation of 13\,\AA. To initiate motion towards the slab, a random velocity component was added to the initial DFT velocities of the ECCO molecule along the $z$‑axis. The velocities of the slab atoms were kept unchanged from the AIMD frames.
The simulations were propagated with a time step of 0.5\,fs for 20\,000 iterations, corresponding to a total simulation time of 10\,ps. An NVE ensemble was employed,\cite{Bigi2024}  as the systems were pre-equilibrated at the target temperature in the DFT stage and the experimental surface-deposition process is intrinsically non-equilibrium.
\subsection{Machine learning}
For machine learning, we employ the foundational messaging passing atomic cluster expansion (MACE)\cite{Batatia2022} model, MACE-MP-0,\cite{Batatia2024} which was originally trained on the Materials Project Trajectory (MPtrj) dataset.\cite{Deng2023a} This dataset contains approximately 1.5\,million configurations, primarily small periodic unit cells representing inorganic crystals with some molecular components.\cite{Batatia2024} Notably, the MPtrj dataset contains limited surface‑chemistry data, motivating the fine‑tuning of MACE‑MP‑0 for improved data efficiency.
Our fine‑tuning approach assumes that knowledge gained from a large and diverse dataset of materials facilitates learning for new systems. Accordingly, the parameters of MACE‑MP‑0 were used to initialise the training of fine‑tuned models.
The model representation comprises of 128 scalar and 128 vectorial components. Fine‑tuning was performed with a learning rate of 0.001 for 100\,epochs to prevent overfitting, re‑initialising the readout layers.\cite{MACE_manual} Training employed a batch size of 16 across 8\,NVIDIA\,A100‑SXM4\,GPUs. A weighting factor of 1:100 between energies and forces was applied, reflecting the greater importance of forces for MD simulations. Five percent of the data was reserved for testing. All other architectural parameters were kept at their default values, matching those of the foundational MACE‑MP‑0 model.\cite{MACE_manual} A fixed random seed of 24 was used for reproducibility.
For comparison, we also trained MACE models from scratch using the same setup as the fine-tuned models, except for an increased training length of 1\,000\,epochs. 
\subsection{MD Analysis}
Trajectory analysis was performed using ASE modules.\cite{Larsen2017,Melix2019e} We further used an automated detection method for adsorption sites and modes. Structure visualizations were rendered using Blender\cite{Blender} via our ASE-Blender interface.\cite{Weiske_Blender2024}
To evaluate the sampling density of the space above the Si(001) surface, a binning approach was conducted using NumPy v2.3.\cite{Harris2020} An $xy$-grid was created and, for each $xy$ bin in the unit cell, the lowest occurring $z$-value of the centre of the cyclooctyne triple bond was stored for the respective set of trajectories. The spacing of the $xy$ bins was set to one thousandth of the unit cell, corresponding to an area of $0.0015\,$\AA$^{2}$ per bin.
\section{Results and Discussion}
\subsection*{Machine learning}
To ensure accurate machine learning interatomic potentials, we analysed the learning and training data distribution using learning curves and dimensionality reduction techniques, respectively.
\begin{figure}[h]
    \centering
    \includegraphics[width=0.6\columnwidth]{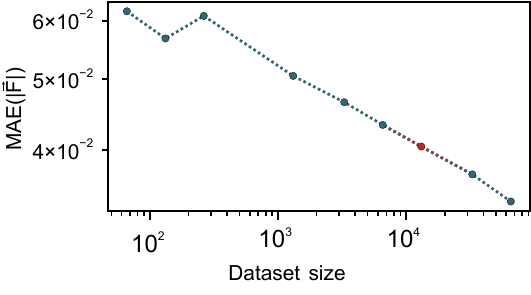}
    \caption{Learning curve of the fine-tuned models showing the mean absolute error (MAE) of forces ($\vec{F}$) plotted against the dataset size in logarithmic scale. The production dataset is marked in red, at 13\,000 data points.}
    \label{fig:learning_curve}
\end{figure}

The learning curves for our fine-tuned models (Figure~\ref{fig:learning_curve}) plot the force mean absolute error (MAE) on an independent test set, with respect to the DFT reference, against the number of training geometries on a log--log scale. The observed decay of the MAE is clearly linear, demonstrating that the fine-tuned model continues to benefit systematically from additional data. 
The energy and force errors for varying training set sizes are presented in detail for all models in Table~S1 of the supplementary information (SI). Fine-tuning generally requires fewer epochs and, consequently, less training time than training models from scratch on the ECCO on Si(001) system (see also Table~\ref{tab:comp_times}). Both models achieve lower errors than the non-fine-tuned foundational model. This is expected as the foundational model does not have knowledge on the data and are also trained on another reference method. Additionally, we find that models trained from scratch on our AIMD data achieve lower errors than the fine-tuned versions from the foundational model ($1.75\times10^{-3}$\,eV and $2.97\times10^{-2}$\,eV/\AA\ for energies and forces, respectively, compared to MAEs of $2.73\times10^{-3}$\,eV and $4.04\times10^{-2}$\,eV/\AA\ for the fine-tuned models). This counter-intuitive result likely stems from differences between the MPtrj data and our target domain. The MPtrj dataset spans a much broader chemical space, containing bonding motifs and structures not directly relevant to our system. Fine-tuning adapts the model to our trajectories, but it begins from parameter values optimized for generalization across the MPtrj dataset. These values turn out to be less suitable for the narrower Si–C–H surface chemistry under study, compared with random initialization when training from scratch. Using a larger learning rate with a decay schedule may help the fine-tuned model escape the local minima associated with the pre-training, potentially achieving errors comparable to the model trained from scratch, but at the cost of losing much of the information gained from the MPtrj dataset. That said, it is important to note that a direct comparison of error metrics, either among different models or against DFT convergence criteria, is not meaningful in this context. 
Nevertheless, the key advantage of fine-tuned models is that they retain knowledge from their pre-training, enabling broader transferability across chemical space. In our case, the strength lies in generalizing to unforeseen configurations not present in the training set that might be seen in a potential MD trajectory, making the model more suitable when generalizability and accurate observables are preferred over minimizing the error over a selection of predefined configurations. To support this, we compared MD simulations of unfunctionalized cyclooctyne at the Si(001) surface using both models, applying the same protocol. Remarkably, the fine-tuned model outperformed the model trained from scratch, resulting in a negligible number of unphysical cyclooctyne structures (see section S4 of the SI for details).

\begin{figure}[h]
    \centering
    \includegraphics[width=0.55\linewidth]{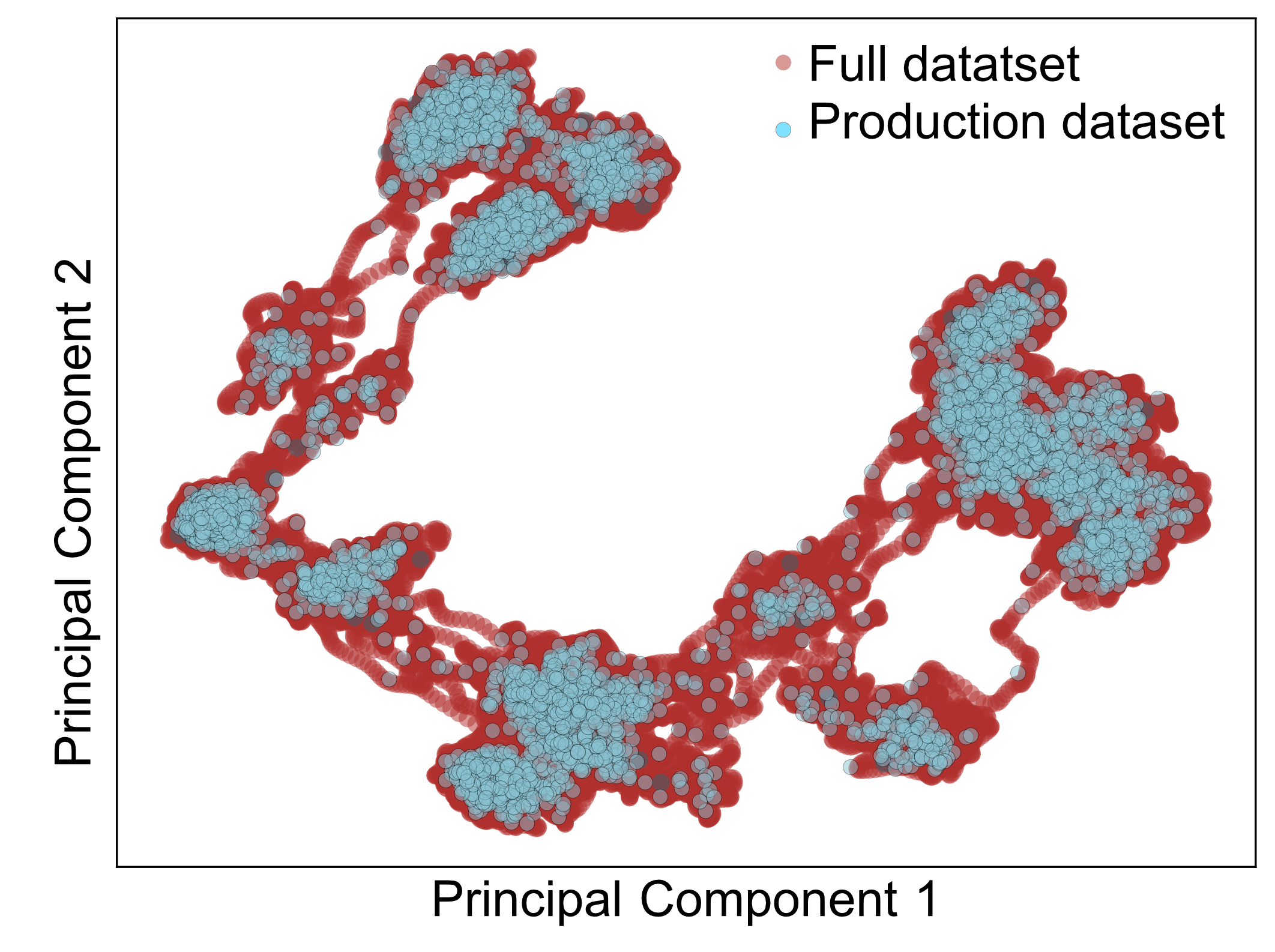}
    \caption{PCA analysis of the full (red) and the production AIMD-datasets (blue). Equivariant features are used for the descriptors as inputs for PCA.}
    \label{fig:PCA-DFT}
\end{figure}
 For final models, we use every 25$^{th}$ AIMD frame, resulting in 13\,000 data points for training. To ensure that these $\sim$13\,000 configurations adequately span relevant reaction pathways and surface environments, we embedded both training and reference AIMD geometries into a low-dimensional manifold using principal component analysis (PCA) based on equivariant geometrical descriptors (Figure~\ref{fig:PCA-DFT}). As shown, both datasets cover approximately the same space, demonstrating the completeness of our training dataset using only every 25$^{th}$ AIMD frame for training. Energy-scaled PCAs for all relevant parts of the dataset are presented in Figure~S6 (SI), indicating that reducing the production dataset size does not significantly reduce the chemical space or energy ranges covered.
\subsection*{ML Driven MD Simulations}
\begin{table}[h]
  \centering
  \caption{Computational time comparison of DFT and machine learning MDs. DFT values are extrapolated from one trajectory. One MD run consists of 20\,000 time steps. All timings in CPUh/GPUh, respectively. Training was performed on 8 GPUs and takes 1.1 h for fine-tuned models and 6.6 h for models trained from scratch.}
  \begin{tabular}{lrrr}
    \hline
    Method & Full MD\textsuperscript{a} & 1 MD Step & 1\,000 MDs  \\
    \hline
    DFT\textsuperscript{b}  &    \num[]{7.1e4}      & \num{3.6e0} & \num[]{7.1e7} \\
    Average ML\textsuperscript{c}  &   \num{1.5e-1}  & \num[]{7.2e-6} & \num[]{1.45e2} \\
    \hline
    \multicolumn{4}{l}{\textsuperscript{a} 20.000 steps, extrapolated from 14,933 steps for DFT-MD.}\\
    \multicolumn{4}{l}{\textsuperscript{b} CPU: 20 Intel Haswell E5-2680v3 = 240 cores total.}\\
    \multicolumn{4}{l}{\textsuperscript{c} GPU: NVIDIA A100-SXM4.}\\
  \end{tabular}
  \label{tab:comp_times}
\end{table}
To assess the role of statistics in MD simulations and to enable meaningful comparison with experiments, we performed 100, 1\,000, and 10\,000 trajectories using machine learning interatomic potentials, in contrast to the 10 trajectories feasible with full DFT-based AIMD. Both machine learning models were tested for their ability to reproduce relevant chemical events in the MD runs. As shown in Table~\ref{tab:comp_times}, using machine learning models results in a dramatic reduction in computational time over AIMD of several orders of magnitude. At the same time, the training time of the fine-tuned model training is only $1.1\,$h on eight GPUs. For the model trained from scratch, increasing the number of epochs by a factor of 10 in comparison to the fine-tuned model, increases the training time by a factor of six to 6.6~h.
\begin{figure}[h]
    \centering
    \includegraphics[width=0.65\columnwidth]{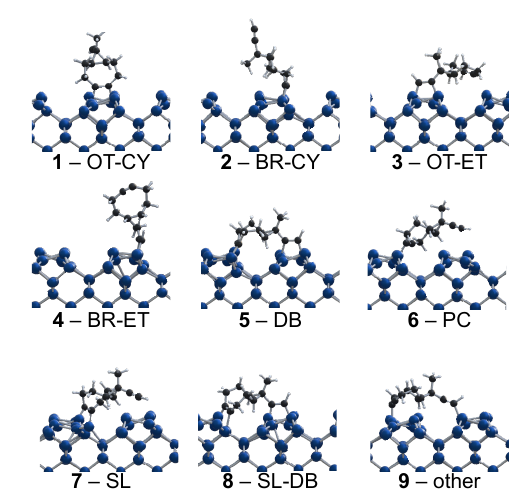}
    \caption{Binding modes as detected using ML-MD and (partially) AIMD ($\mathbf{1-9}$). Si atoms blue, C black and H white. ECCO binds either via the triple bond of the cyclooctyne (CY) or ethinyl (ET) group. The Si(001) surface offers two distinct binding sites on-top (OT) and bridge (BR). As each ECCO can bind via the two functional groups, also doubly bound ECCO molecules can be observed (DB). During reactions of ECCO with the surface, ECCO molecules that bind to a single Si atom can be observed (precursor states, PC). When Si atoms of the second Si atom layer are involved in bonding, structures are labeled as sublayer (SL). These binding modes can also occur with doubly bound ECCO molecules(sublayer-double, SL-DB).}
    \label{fig:structures}
\end{figure}

To analyse whether improved statistics lead to larger sampling of configuration space during dynamics and new structures not observed in the 10 DFT-based AIMD, we analyse representative adsorption structures. These are illustrated in Figure~\ref{fig:structures}. Nine representative adsorption structures arise in our molecular gun simulations: \textbf{1}--on-top cyclooctyne (OT-CY), \textbf{2}--bridge cyclooctyne (BR-CY), \textbf{3}--on-top ethinyl (OT-ET), \textbf{4}--bridge ethinyl (BR-ET), \textbf{5}--double (DB), \textbf{6}--precursor (PC), \textbf{7}--sublayer (SL), \textbf{8}--sublayer double (SL-DB), and \textbf{9}--other.

Configurations \textbf{1}--\textbf{4} involve the molecule spanning two adjacent surface atoms, either via the cyclooctyne ring's triple bond (CY; \textbf{1} and \textbf{2}) or the ethinyl group (ET; \textbf{3} and \textbf{4}). The two surface atoms can be on the same Si dimer (on-top, OT; \textbf{1} and \textbf{3}) or on neighboring dimers (bridge, BR; \textbf{2} and \textbf{4}). Configuration \textbf{5} comprises states where both triple bonds react with the surface to form doubly bonded ECCO (DB). The precursor state (PC; \textbf{6}) describes ECCO datively bound to a single Si atom. This state is observed in DFT data, where it is an important reaction intermediate.\cite{Pieck2021} In sublayer (SL; \textbf{7}) and sublayer-double (SL-DB; \textbf{8}) structures, ECCO binds to a Si atom beneath the top layer, either singly or doubly. All other configurations (mainly both triple bonds in a datively bonded state) are grouped as \textit{other} (\textbf{9}).

\begin{figure}[h]
   \centering
   \includegraphics[width=0.65\columnwidth]{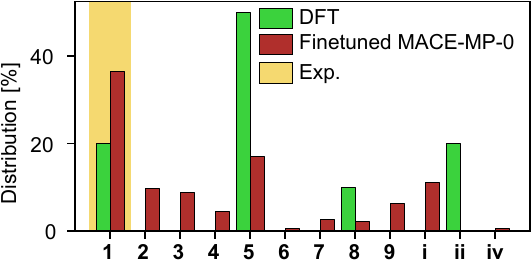}
   \caption{Distribution of the binding modes at the end of each MD simulation for the fine-tuned MACE-MP-0 model and the the AIMD reference. The binding mode distribution of the fine-tuned model is given in red and the AIMD reference in green. The experimentally observed binding mode is highlighted in yellow.}
   \label{fig:barplots}
\end{figure}
Figure~\ref{fig:barplots} shows the distribution of final ECCO adsorption sites on Si(001) for the fine-tuned model and the AIMD reference. The distributions obtained for the other discussed models are shown in Figure~S1. The main experimental adsorption mode corresponds to the on-top cyclooctyne (\textbf{1}), shown in yellow.\cite{Langer2019} The 10 DFT-based AIMDs fail to capture the experimentally dominant motif \textbf{1}, while instead over-emphasizing the doubly bonded mode \textbf{5}. The 1\,000 trajectories obtained using the fine-tuned  MACE-MP-0 model are able to capture the experimental motif much better, recovering the statistics of the adsorption. Additionally, a heap of other final states are observed, which don't occur as final states in the DFT trajectories resulting in a more realistic distribution.

Not all MD trajectories yield covalent binding -- non-productive outcomes are labelled: intact but floating ECCO (\textbf{i}), H-abstraction (\textbf{ii}), broken C--C bonds (\textbf{iii}, see Figure~S5, SI), and ``explosions'' (\textbf{iv}). As shown in Figure~S1a, the fine-tuned model best matches experiment. From-scratch models produce a high fraction ($>30\,\%$) of desorbed intact ECCO (\textbf{i}), while MACE-MP-0 often breaks the C--C bond ($\sim20\,\%$; \textbf{iii}). The tendency of the MACE-MP-0 and from-scratch model yield unphysical or unproductive events, despite low MAEs, underscores that the fine-tuned model generalizes better and predicts realistic dynamics at scale.

Figure~S1b illustrates convergence of binding mode populations with increasing number of MD trajectories. 
Moving from 100 to 1\,000 machine learning based MD trajectories, there is a marked shift in the observed binding mode distribution and covered configurational space, as supported by PCA descriptor plots (see Figure~S7). The difference between 1\,000 and 10\,000 trajectories, however, is minor, indicating statistical convergence is reached around 1\,000 independent simulations. The precise proportions of binding modes for 1\,000 and 10\,000 runs are summarized in Table~S2.
\begin{figure}[h]
    \centering
    \includegraphics[width=0.65\columnwidth]{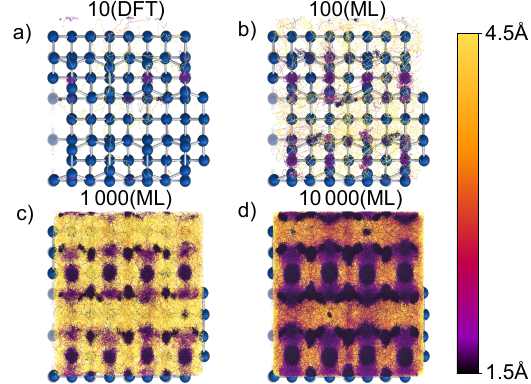}
    \caption{Representation of the sampling (top view). Utilizing a binning approach on the center of the cyclooctyne triple bond over all simulation runs for (a) 10 DFT, (b) 100, (c) 1\,000, and (d) 10\,000 trajectories obtained using the fine-tuned MACE-MP-0 model. The bins are created with a size of $0.0015\,\text{\AA}^2$ and colored corresponding to the lowest occurring z-value (surface normal). The height information is relative to the topmost surface atoms.}
    \label{fig:main_sampling_surface}
\end{figure}

Examining surface sampling, Figure~\ref{fig:main_sampling_surface} displays the minimum distance between the cyclooctyne triple bond and the surface for all trajectories. DFT-based trajectories reach only a limited set of surface sites, reflecting the small sample. With 100 machine learning-driven simulations using the fine-tuned model, coverage expands, but many regions remain unsampled. For both 1\,000 and 10\,000 trajectories, all surface regions are visited, supporting the conclusion from Figure~S1b that 1\,000 runs suffice for statistical convergence. (See Figure~S3 for mode-colored sampling, and Figure~S4 for a side view.)
\subsection*{Preferred Binding Mode}
The statistical convergence of binding mode distributions demonstrates that AIMD simulations based on only 10 DFT trajectories are insufficient to provide a realistic depiction of ECCO adsorption behavior. Most notably, the most prominent double-adsorption structure (\textbf{5}) appears primarily because the PBE functional underestimates the ethinyl reaction barrier by approximately $0.1$ to $0.2$\,eV, as previously shown by Pecher and Tonner-Zech.\cite{pecher_bond_2018} This underestimation, and the consequent increased likelihood for the ethinyl group to react with the surface, are therefore also inherited by our ML models.
\section{Conclusion}
We have demonstrated that large-scale molecular dynamics sampling is essential to accurately reproduce experimental adsorption statistics for large molecules adsorbing on a surface at the example of ethinyl-functionalized cyclooctyne on Si(001). By leveraging equivariant message-passing neural network potentials, comparing models trained from scratch and fine-tuned models based on parameters of foundational models, we achieved over $10^{3}$-fold speed-ups compared to conventional DFT-based molecular dynamics, enabling 10$^{3}$–10$^{4}$ trajectories to be run with moderate computational resources. Fine-tuning on a few thousand AIMD snapshots is critical to adapt foundational models for specific surface chemistry: without this step, important adsorption modes are missing in subsequent machine learning-driven MD. Moreover, fine-tuning requires only a handful of epochs, reducing both training times and data requirements, while also minimizing the risk of catastrophic forgetting compared to training from scratch. 

Our high-throughput based MDs uncover numerous new final states, such as mixed on-top/sublayer motifs, which are rare or completely absent in the few DFT-based runs. This comprehensive sampling shifts theoretical predictions towards the experimentally dominant on-top adsorption mode. Nevertheless, certain DFT-induced biases persist, in particular the tendency for the double-adsorption motif to appear due to the PBE underestimation of the ethinyl reaction barrier, which is inherited by the machine learning models. As the number of trajectories increases, surface sampling rapidly improves: at 100 runs, significant regions remain unsampled; at 1\,000, all key binding motifs are visited; at 10\,000, near-complete coverage of the surface is achieved.

The statistical convergence of site populations between 1\,000 and 10\,000 trajectories is minor, clearly indicating that poor statistical sampling -- rather than deficiencies in \textit{ab-initio} theory -- explains discrepancies between previous AIMD and experimental studies. Future work could further improve accuracy by employing $\Delta$-learning, fine-tuning against higher-level quantum data, or selectively incorporating experimental observables to address residual DFT errors. Overall, our “machine-learning molecular gun” workflow provides a robust and scalable means to connect atomistic mechanisms with ensemble-level experiments, thereby guiding the rational design of surface-functionalized semiconductor devices. Due to the small fine-tuning required, this model can easily be extended to other surfaces and adsorbates with the potential to significantly increase the impact of modelling in computational surface dynamics.
\section*{Author contributions}
Hendrik Weiske: Data Curation, Formal Analysis, Investigation, Methodology, Software, Validation, Visualization, Writing - Review \& Editing;
Rhyan Barett: ML Methodology, Writing - Editing;
Ralf Tonner-Zech: Conceptualization, Funding Acquisition, Project Administration, Resources, Supervision, Writing - Review \& Editing;
Patrick Melix: Data Curation, Formal Analysis, Methodology, Project Administration, Software, Supervision, Writing - Original Draft Preparation, Review \& Editing;
Julia Westermayr: Funding Acquisition, Project Administration, Resources, Supervision, Writing - Original Draft Preparation, Review \& Editing.

During manuscript preparation, the authors used the Perplexity AI writing assistant to refine language, improve clarity, and strengthen organization. AI suggestions were used to polish text, ensure consistency between sections, and enhance readability. All scientific data analysis, methodological development, interpretation, and content decisions were performed by the authors, who take full responsibility for the manuscript’s originality and accuracy.
\section*{Conflicts of interest}
There are no conflicts to declare.
\section*{Data availability}
The data used to train ML models is freely available from reference \citenum{Pieck2021a}.
\section*{Code availability}
The code used to conduct molecular dynamics with the MACE model is freely available via the ASE package.\cite{Larsen2017} Instructions to train MACE models and foundational models using fine-tuning processes can be found in \citenum{MACE_manual}. All scripts and outputs (excluding ML-MD trajectories due to size) produced in this project are available in the published raw data on Zenodo.\cite{Weiske2025_data} 
\section*{Acknowledgements}
The authors thank L. Yang, S. Schumann, T. Oestereich, and D. Bitterlich for their help in preprocessing the data and setting up the MACE fine-tuning procedure. We thank Dr. Fabian Pieck for providing the AIMD trajectories and discussion. This work was supported by the Deutsche Forschungsgemeinschaft (DFG) via SFB1083 "Structure and Dynamics of Internal Interfaces" and GRK 2721 “Hydrogen Isotopes 1,2,3H”.  Computations for this work were done using resources of ZIH Dresden, NHR-PC2 Paderborn, CSC-Goethe Frankfurt, and the Leipzig University Computing Center. 


\balance

\providecommand*{\mcitethebibliography}{\thebibliography}
\csname @ifundefined\endcsname{endmcitethebibliography}
{\let\endmcitethebibliography\endthebibliography}{}


\begin{mcitethebibliography}{69}
\providecommand*{\natexlab}[1]{#1}
\providecommand*{\mciteSetBstSublistMode}[1]{}
\providecommand*{\mciteSetBstMaxWidthForm}[2]{}
\providecommand*{\mciteBstWouldAddEndPuncttrue}
  {\def\EndOfBibitem{\unskip.}}
\providecommand*{\mciteBstWouldAddEndPunctfalse}
  {\let\EndOfBibitem\relax}
\providecommand*{\mciteSetBstMidEndSepPunct}[3]{}
\providecommand*{\mciteSetBstSublistLabelBeginEnd}[3]{}
\providecommand*{\EndOfBibitem}{}
\mciteSetBstSublistMode{f}
\mciteSetBstMaxWidthForm{subitem}
{(\emph{\alph{mcitesubitemcount}})}
\mciteSetBstSublistLabelBeginEnd{\mcitemaxwidthsubitemform\space}
{\relax}{\relax}

\bibitem[Wolkow(1999)]{Wolkow1999}
R.~A. Wolkow, \emph{Annual Review of Physical Chemistry}, 1999, \textbf{50},
  413--441\relax
\mciteBstWouldAddEndPuncttrue
\mciteSetBstMidEndSepPunct{\mcitedefaultmidpunct}
{\mcitedefaultendpunct}{\mcitedefaultseppunct}\relax
\EndOfBibitem
\bibitem[Teplyakov and Bent(2013)]{Teplyakov2013a}
A.~V. Teplyakov and S.~F. Bent, \emph{Journal of Vacuum Science \& Technology
  A: Vacuum, Surfaces, and Films}, 2013, \textbf{31}, 050810\relax
\mciteBstWouldAddEndPuncttrue
\mciteSetBstMidEndSepPunct{\mcitedefaultmidpunct}
{\mcitedefaultendpunct}{\mcitedefaultseppunct}\relax
\EndOfBibitem
\bibitem[Bent(2002)]{Bent2002}
S.~F. Bent, \emph{Surface Science}, 2002, \textbf{500}, 879--903\relax
\mciteBstWouldAddEndPuncttrue
\mciteSetBstMidEndSepPunct{\mcitedefaultmidpunct}
{\mcitedefaultendpunct}{\mcitedefaultseppunct}\relax
\EndOfBibitem
\bibitem[Kachian \emph{et~al.}(2010)Kachian, Wong, and Bent]{Kachian2010}
J.~S. Kachian, K.~T. Wong and S.~F. Bent, \emph{Accounts of Chemical Research},
  2010, \textbf{43}, 346--355\relax
\mciteBstWouldAddEndPuncttrue
\mciteSetBstMidEndSepPunct{\mcitedefaultmidpunct}
{\mcitedefaultendpunct}{\mcitedefaultseppunct}\relax
\EndOfBibitem
\bibitem[Kolb \emph{et~al.}(2001)Kolb, Finn, and Sharpless]{Kolb2001}
H.~C. Kolb, M.~G. Finn and K.~B. Sharpless, \emph{Angewandte Chemie
  International Edition}, 2001, \textbf{40}, 2004--2021\relax
\mciteBstWouldAddEndPuncttrue
\mciteSetBstMidEndSepPunct{\mcitedefaultmidpunct}
{\mcitedefaultendpunct}{\mcitedefaultseppunct}\relax
\EndOfBibitem
\bibitem[Münster \emph{et~al.}(2016)Münster, Nikodemiak, and
  Koert]{Munster2016}
N.~Münster, P.~Nikodemiak and U.~Koert, \emph{Organic Letters}, 2016,
  \textbf{18}, 4296--4299\relax
\mciteBstWouldAddEndPuncttrue
\mciteSetBstMidEndSepPunct{\mcitedefaultmidpunct}
{\mcitedefaultendpunct}{\mcitedefaultseppunct}\relax
\EndOfBibitem
\bibitem[Reutzel \emph{et~al.}(2016)Reutzel, Münster, Lipponer, Länger,
  Höfer, Koert, and Dürr]{Reutzel2016}
M.~Reutzel, N.~Münster, M.~A. Lipponer, C.~Länger, U.~Höfer, U.~Koert and
  M.~Dürr, \emph{Journal of Physical Chemistry C}, 2016, \textbf{120},
  26284--26289\relax
\mciteBstWouldAddEndPuncttrue
\mciteSetBstMidEndSepPunct{\mcitedefaultmidpunct}
{\mcitedefaultendpunct}{\mcitedefaultseppunct}\relax
\EndOfBibitem
\bibitem[Glaser \emph{et~al.}(2021)Glaser, Meinecke, Freund, Länger, Luy,
  Tonner, Koert, and Dürr]{Glaser2021}
T.~Glaser, J.~Meinecke, L.~Freund, C.~Länger, J.-N. Luy, R.~Tonner, U.~Koert
  and M.~Dürr, \emph{Chemistry – A European Journal}, 2021, \textbf{27},
  8082--8087\relax
\mciteBstWouldAddEndPuncttrue
\mciteSetBstMidEndSepPunct{\mcitedefaultmidpunct}
{\mcitedefaultendpunct}{\mcitedefaultseppunct}\relax
\EndOfBibitem
\bibitem[Glaser \emph{et~al.}(2021)Glaser, Meinecke, Länger, Luy, Tonner,
  Koert, and Dürr]{Glaser2021a}
T.~Glaser, J.~Meinecke, C.~Länger, J.-N. Luy, R.~Tonner, U.~Koert and
  M.~Dürr, \emph{ChemPhysChem}, 2021, \textbf{22}, 404--409\relax
\mciteBstWouldAddEndPuncttrue
\mciteSetBstMidEndSepPunct{\mcitedefaultmidpunct}
{\mcitedefaultendpunct}{\mcitedefaultseppunct}\relax
\EndOfBibitem
\bibitem[Nalaoh \emph{et~al.}(2025)Nalaoh, Clark, Arroyo-Currás, and
  Jenkins]{Nalaoh2025}
P.~Nalaoh, V.~Clark, N.~Arroyo-Currás and D.~M. Jenkins, \emph{ACS Sensors},
  2025\relax
\mciteBstWouldAddEndPuncttrue
\mciteSetBstMidEndSepPunct{\mcitedefaultmidpunct}
{\mcitedefaultendpunct}{\mcitedefaultseppunct}\relax
\EndOfBibitem
\bibitem[Dürr \emph{et~al.}(2025)Dürr, Höfer, Koert, and
  Tonner-Zech]{Durr2025}
M.~Dürr, U.~Höfer, U.~Koert and R.~Tonner-Zech, \emph{Accounts of Chemical
  Research}, 2025, \textbf{58}, 2454--2465\relax
\mciteBstWouldAddEndPuncttrue
\mciteSetBstMidEndSepPunct{\mcitedefaultmidpunct}
{\mcitedefaultendpunct}{\mcitedefaultseppunct}\relax
\EndOfBibitem
\bibitem[Mette \emph{et~al.}(2013)Mette, Dürr, Bartholomäus, Koert, and
  Höfer]{Mette2013}
G.~Mette, M.~Dürr, R.~Bartholomäus, U.~Koert and U.~Höfer, \emph{Chemical
  Physics Letters}, 2013, \textbf{556}, 70--76\relax
\mciteBstWouldAddEndPuncttrue
\mciteSetBstMidEndSepPunct{\mcitedefaultmidpunct}
{\mcitedefaultendpunct}{\mcitedefaultseppunct}\relax
\EndOfBibitem
\bibitem[Pecher \emph{et~al.}(2017)Pecher, Schmidt, and Tonner]{Pecher2017c}
L.~Pecher, S.~Schmidt and R.~Tonner, \emph{Journal of Physical Chemistry C},
  2017, \textbf{121}, 26840--26850\relax
\mciteBstWouldAddEndPuncttrue
\mciteSetBstMidEndSepPunct{\mcitedefaultmidpunct}
{\mcitedefaultendpunct}{\mcitedefaultseppunct}\relax
\EndOfBibitem
\bibitem[Pecher \emph{et~al.}(2017)Pecher, Schober, and Tonner]{Pecher2017a}
L.~Pecher, C.~Schober and R.~Tonner, \emph{Chemistry – A European Journal},
  2017, \textbf{23}, 5459--5466\relax
\mciteBstWouldAddEndPuncttrue
\mciteSetBstMidEndSepPunct{\mcitedefaultmidpunct}
{\mcitedefaultendpunct}{\mcitedefaultseppunct}\relax
\EndOfBibitem
\bibitem[Pecher \emph{et~al.}(2018)Pecher, Schmidt, and Tonner]{Pecher2018b}
L.~Pecher, S.~Schmidt and R.~Tonner, \emph{Beilstein Journal of Organic
  Chemistry}, 2018, \textbf{14}, 2715--2721\relax
\mciteBstWouldAddEndPuncttrue
\mciteSetBstMidEndSepPunct{\mcitedefaultmidpunct}
{\mcitedefaultendpunct}{\mcitedefaultseppunct}\relax
\EndOfBibitem
\bibitem[Pecher and Tonner(2018)]{Pecher2018}
L.~Pecher and R.~Tonner, \emph{Theoretical Chemistry Accounts}, 2018,
  \textbf{137}, 48\relax
\mciteBstWouldAddEndPuncttrue
\mciteSetBstMidEndSepPunct{\mcitedefaultmidpunct}
{\mcitedefaultendpunct}{\mcitedefaultseppunct}\relax
\EndOfBibitem
\bibitem[Glaser \emph{et~al.}(2020)Glaser, Länger, Heep, Meinecke, Silly,
  Koert, and Dürr]{Glaser2020}
T.~Glaser, C.~Länger, J.~Heep, J.~Meinecke, M.~G. Silly, U.~Koert and
  M.~Dürr, \emph{The Journal of Physical Chemistry C}, 2020, \textbf{124},
  22619--22624\relax
\mciteBstWouldAddEndPuncttrue
\mciteSetBstMidEndSepPunct{\mcitedefaultmidpunct}
{\mcitedefaultendpunct}{\mcitedefaultseppunct}\relax
\EndOfBibitem
\bibitem[Glaser \emph{et~al.}(2024)Glaser, Peters, Scharf, Koert, and
  Dürr]{Glaser2024}
T.~Glaser, J.~A. Peters, D.~Scharf, U.~Koert and M.~Dürr, \emph{Chemistry of
  Materials}, 2024, \textbf{36}, 561--566\relax
\mciteBstWouldAddEndPuncttrue
\mciteSetBstMidEndSepPunct{\mcitedefaultmidpunct}
{\mcitedefaultendpunct}{\mcitedefaultseppunct}\relax
\EndOfBibitem
\bibitem[Länger \emph{et~al.}(2019)Länger, Heep, Nikodemiak, Bohamud,
  Kirsten, Höfer, Koert, and Dürr]{Langer2019}
C.~Länger, J.~Heep, P.~Nikodemiak, T.~Bohamud, P.~Kirsten, U.~Höfer, U.~Koert
  and M.~Dürr, \emph{Journal of Physics: Condensed Matter}, 2019, \textbf{31},
  034001\relax
\mciteBstWouldAddEndPuncttrue
\mciteSetBstMidEndSepPunct{\mcitedefaultmidpunct}
{\mcitedefaultendpunct}{\mcitedefaultseppunct}\relax
\EndOfBibitem
\bibitem[Maurer \emph{et~al.}(2016)Maurer, Ruiz, Camarillo-Cisneros, Liu,
  Ferri, Reuter, and Tkatchenko]{Maurer2016}
R.~J. Maurer, V.~G. Ruiz, J.~Camarillo-Cisneros, W.~Liu, N.~Ferri, K.~Reuter
  and A.~Tkatchenko, \emph{Progress in Surface Science}, 2016, \textbf{91},
  72--100\relax
\mciteBstWouldAddEndPuncttrue
\mciteSetBstMidEndSepPunct{\mcitedefaultmidpunct}
{\mcitedefaultendpunct}{\mcitedefaultseppunct}\relax
\EndOfBibitem
\bibitem[Maurer \emph{et~al.}(2019)Maurer, Freysoldt, Reilly, Brandenburg,
  Hofmann, Björkman, Lebègue, and Tkatchenko]{Maurer2019}
R.~J. Maurer, C.~Freysoldt, A.~M. Reilly, J.~G. Brandenburg, O.~T. Hofmann,
  T.~Björkman, S.~Lebègue and A.~Tkatchenko, \emph{Annual Review of Materials
  Research}, 2019, \textbf{49}, 1--30\relax
\mciteBstWouldAddEndPuncttrue
\mciteSetBstMidEndSepPunct{\mcitedefaultmidpunct}
{\mcitedefaultendpunct}{\mcitedefaultseppunct}\relax
\EndOfBibitem
\bibitem[Van~Duin \emph{et~al.}(2001)Van~Duin, Dasgupta, Lorant, and
  Goddard]{VanDuin2001}
A.~C.~T. Van~Duin, S.~Dasgupta, F.~Lorant and W.~A. Goddard, \emph{The Journal
  of Physical Chemistry A}, 2001, \textbf{105}, 9396--9409\relax
\mciteBstWouldAddEndPuncttrue
\mciteSetBstMidEndSepPunct{\mcitedefaultmidpunct}
{\mcitedefaultendpunct}{\mcitedefaultseppunct}\relax
\EndOfBibitem
\bibitem[Senftle \emph{et~al.}(2016)Senftle, Hong, Islam, Kylasa, Zheng, Shin,
  Junkermeier, Engel-Herbert, Janik, Aktulga, Verstraelen, Grama, and
  Van~Duin]{Senftle2016}
T.~P. Senftle, S.~Hong, M.~M. Islam, S.~B. Kylasa, Y.~Zheng, Y.~K. Shin,
  C.~Junkermeier, R.~Engel-Herbert, M.~J. Janik, H.~M. Aktulga, T.~Verstraelen,
  A.~Grama and A.~C.~T. Van~Duin, \emph{npj Computational Materials}, 2016,
  \textbf{2}, 15011\relax
\mciteBstWouldAddEndPuncttrue
\mciteSetBstMidEndSepPunct{\mcitedefaultmidpunct}
{\mcitedefaultendpunct}{\mcitedefaultseppunct}\relax
\EndOfBibitem
\bibitem[Hu \emph{et~al.}(2017)Hu, Schuster, and Schulz]{Hu2017b}
X.~Hu, J.~Schuster and S.~E. Schulz, \emph{The Journal of Physical Chemistry
  C}, 2017, \textbf{121}, 28077--28089\relax
\mciteBstWouldAddEndPuncttrue
\mciteSetBstMidEndSepPunct{\mcitedefaultmidpunct}
{\mcitedefaultendpunct}{\mcitedefaultseppunct}\relax
\EndOfBibitem
\bibitem[Zhu \emph{et~al.}(2020)Zhu, Gong, Han, Zhang, and Van~Duin]{Zhu2020}
W.~Zhu, H.~Gong, Y.~Han, M.~Zhang and A.~C.~T. Van~Duin, \emph{The Journal of
  Physical Chemistry C}, 2020, \textbf{124}, 12512--12520\relax
\mciteBstWouldAddEndPuncttrue
\mciteSetBstMidEndSepPunct{\mcitedefaultmidpunct}
{\mcitedefaultendpunct}{\mcitedefaultseppunct}\relax
\EndOfBibitem
\bibitem[Wen \emph{et~al.}(2017)Wen, Ma, Zhang, Van~Duin, and Lu]{Wen2017}
J.~Wen, T.~Ma, W.~Zhang, A.~C.~T. Van~Duin and X.~Lu, \emph{The Journal of
  Physical Chemistry A}, 2017, \textbf{121}, 587--594\relax
\mciteBstWouldAddEndPuncttrue
\mciteSetBstMidEndSepPunct{\mcitedefaultmidpunct}
{\mcitedefaultendpunct}{\mcitedefaultseppunct}\relax
\EndOfBibitem
\bibitem[Nayir \emph{et~al.}(2019)Nayir, Van~Duin, and Erkoc]{Nayir2019}
N.~Nayir, A.~C.~T. Van~Duin and S.~Erkoc, \emph{The Journal of Physical
  Chemistry A}, 2019, \textbf{123}, 4303--4313\relax
\mciteBstWouldAddEndPuncttrue
\mciteSetBstMidEndSepPunct{\mcitedefaultmidpunct}
{\mcitedefaultendpunct}{\mcitedefaultseppunct}\relax
\EndOfBibitem
\bibitem[Iftimie \emph{et~al.}(2005)Iftimie, Minary, and
  Tuckerman]{Iftimie2005}
R.~Iftimie, P.~Minary and M.~E. Tuckerman, \emph{Proceedings of the National
  Academy of Sciences}, 2005, \textbf{102}, 6654--6659\relax
\mciteBstWouldAddEndPuncttrue
\mciteSetBstMidEndSepPunct{\mcitedefaultmidpunct}
{\mcitedefaultendpunct}{\mcitedefaultseppunct}\relax
\EndOfBibitem
\bibitem[Radeke and Carter(1997)]{Radeke1997}
M.~R. Radeke and E.~A. Carter, \emph{Annual Review of Physical Chemistry},
  1997, \textbf{48}, 243--270\relax
\mciteBstWouldAddEndPuncttrue
\mciteSetBstMidEndSepPunct{\mcitedefaultmidpunct}
{\mcitedefaultendpunct}{\mcitedefaultseppunct}\relax
\EndOfBibitem
\bibitem[Marx and Hutter(2000)]{Grotendorst2000a}
D.~Marx and J.~Hutter, \emph{Modern methods and algorithms of quantum
  chemistry.}, NIC, Jülich, 2000, pp. 301--449\relax
\mciteBstWouldAddEndPuncttrue
\mciteSetBstMidEndSepPunct{\mcitedefaultmidpunct}
{\mcitedefaultendpunct}{\mcitedefaultseppunct}\relax
\EndOfBibitem
\bibitem[Groß(2023)]{Gross2023}
A.~Groß, \emph{Current Opinion in Electrochemistry}, 2023, \textbf{40},
  101345\relax
\mciteBstWouldAddEndPuncttrue
\mciteSetBstMidEndSepPunct{\mcitedefaultmidpunct}
{\mcitedefaultendpunct}{\mcitedefaultseppunct}\relax
\EndOfBibitem
\bibitem[Pieck and Tonner-Zech(2021)]{Pieck2021}
F.~Pieck and R.~Tonner-Zech, \emph{Molecules}, 2021, \textbf{26}, 6653\relax
\mciteBstWouldAddEndPuncttrue
\mciteSetBstMidEndSepPunct{\mcitedefaultmidpunct}
{\mcitedefaultendpunct}{\mcitedefaultseppunct}\relax
\EndOfBibitem
\bibitem[Batatia \emph{et~al.}(2022)Batatia, Kovacs, Simm, Ortner, and
  Csanyi]{Batatia2022}
I.~Batatia, D.~P. Kovacs, G.~Simm, C.~Ortner and G.~Csanyi, Advances in
  {Neural} {Information} {Processing} {Systems}, 2022, pp. 11423--11436\relax
\mciteBstWouldAddEndPuncttrue
\mciteSetBstMidEndSepPunct{\mcitedefaultmidpunct}
{\mcitedefaultendpunct}{\mcitedefaultseppunct}\relax
\EndOfBibitem
\bibitem[Batatia \emph{et~al.}(2024)Batatia, Benner, Chiang, Elena, Kovács,
  Riebesell, Advincula, Asta, Avaylon, Baldwin, Berger, Bernstein, Bhowmik,
  Blau, Cărare, Darby, De, Della~Pia, Deringer, Elijošius, El-Machachi,
  Falcioni, Fako, Ferrari, Genreith-Schriever, George, Goodall, Grey, Grigorev,
  Han, Handley, Heenen, Hermansson, Holm, Jaafar, Hofmann, Jakob, Jung, Kapil,
  Kaplan, Karimitari, Kermode, Kroupa, Kullgren, Kuner, Kuryla, Liepuoniute,
  Margraf, Magdău, Michaelides, Moore, Naik, Niblett, Norwood, O'Neill,
  Ortner, Persson, Reuter, Rosen, Schaaf, Schran, Shi, Sivonxay, Stenczel,
  Svahn, Sutton, Swinburne, Tilly, van~der Oord, Varga-Umbrich, Vegge,
  Vondrák, Wang, Witt, Zills, and Csányi]{Batatia2024}
I.~Batatia, P.~Benner, Y.~Chiang, A.~M. Elena, D.~P. Kovács, J.~Riebesell,
  X.~R. Advincula, M.~Asta, M.~Avaylon, W.~J. Baldwin, F.~Berger, N.~Bernstein,
  A.~Bhowmik, S.~M. Blau, V.~Cărare, J.~P. Darby, S.~De, F.~Della~Pia, V.~L.
  Deringer, R.~Elijošius, Z.~El-Machachi, F.~Falcioni, E.~Fako, A.~C. Ferrari,
  A.~Genreith-Schriever, J.~George, R.~E.~A. Goodall, C.~P. Grey, P.~Grigorev,
  S.~Han, W.~Handley, H.~H. Heenen, K.~Hermansson, C.~Holm, J.~Jaafar,
  S.~Hofmann, K.~S. Jakob, H.~Jung, V.~Kapil, A.~D. Kaplan, N.~Karimitari,
  J.~R. Kermode, N.~Kroupa, J.~Kullgren, M.~C. Kuner, D.~Kuryla,
  G.~Liepuoniute, J.~T. Margraf, I.-B. Magdău, A.~Michaelides, J.~H. Moore,
  A.~A. Naik, S.~P. Niblett, S.~W. Norwood, N.~O'Neill, C.~Ortner, K.~A.
  Persson, K.~Reuter, A.~S. Rosen, L.~L. Schaaf, C.~Schran, B.~X. Shi,
  E.~Sivonxay, T.~K. Stenczel, V.~Svahn, C.~Sutton, T.~D. Swinburne, J.~Tilly,
  C.~van~der Oord, E.~Varga-Umbrich, T.~Vegge, M.~Vondrák, Y.~Wang, W.~C.
  Witt, F.~Zills and G.~Csányi, \emph{A foundation model for atomistic
  materials chemistry}, 2024, \url{https://arxiv.org/abs/2401.00096}, Version
  Number: 2\relax
\mciteBstWouldAddEndPuncttrue
\mciteSetBstMidEndSepPunct{\mcitedefaultmidpunct}
{\mcitedefaultendpunct}{\mcitedefaultseppunct}\relax
\EndOfBibitem
\bibitem[Stark \emph{et~al.}(2024)Stark, Van Der~Oord, Batatia, Zhang, Jiang,
  Csányi, and Maurer]{Stark2024}
W.~G. Stark, C.~Van Der~Oord, I.~Batatia, Y.~Zhang, B.~Jiang, G.~Csányi and
  R.~J. Maurer, \emph{Machine Learning: Science and Technology}, 2024,
  \textbf{5}, 030501\relax
\mciteBstWouldAddEndPuncttrue
\mciteSetBstMidEndSepPunct{\mcitedefaultmidpunct}
{\mcitedefaultendpunct}{\mcitedefaultseppunct}\relax
\EndOfBibitem
\bibitem[Giese \emph{et~al.}(2025)Giese, Zeng, and York]{Giese2025}
T.~J. Giese, J.~Zeng and D.~M. York, \emph{The Journal of Physical Chemistry
  B}, 2025, \textbf{129}, 5477--5490\relax
\mciteBstWouldAddEndPuncttrue
\mciteSetBstMidEndSepPunct{\mcitedefaultmidpunct}
{\mcitedefaultendpunct}{\mcitedefaultseppunct}\relax
\EndOfBibitem
\bibitem[Shiota \emph{et~al.}(2024)Shiota, Ishihara, and Mizukami]{Shiota2024}
T.~Shiota, K.~Ishihara and W.~Mizukami, \emph{Digital Discovery}, 2024,
  \textbf{3}, 1714--1728\relax
\mciteBstWouldAddEndPuncttrue
\mciteSetBstMidEndSepPunct{\mcitedefaultmidpunct}
{\mcitedefaultendpunct}{\mcitedefaultseppunct}\relax
\EndOfBibitem
\bibitem[Schaaf \emph{et~al.}(2024)Schaaf, Rhodes, Zick, Pugh, Hilliard,
  Sharma, Wade, Milner, Csanyi, and Forse]{Schaaf2024}
L.~L. Schaaf, B.~Rhodes, M.~Zick, S.~Pugh, J.~Hilliard, S.~Sharma, C.~Wade,
  P.~Milner, G.~Csanyi and A.~Forse, {AI} for {Accelerated} {Materials}
  {Design}, 2024\relax
\mciteBstWouldAddEndPuncttrue
\mciteSetBstMidEndSepPunct{\mcitedefaultmidpunct}
{\mcitedefaultendpunct}{\mcitedefaultseppunct}\relax
\EndOfBibitem
\bibitem[Sauer \emph{et~al.}(2025)Sauer, Lyngby, and Thygesen]{Sauer2025}
M.~O. Sauer, P.~M. Lyngby and K.~S. Thygesen, \emph{Physical Review Materials},
  2025, \textbf{9}, 074007\relax
\mciteBstWouldAddEndPuncttrue
\mciteSetBstMidEndSepPunct{\mcitedefaultmidpunct}
{\mcitedefaultendpunct}{\mcitedefaultseppunct}\relax
\EndOfBibitem
\bibitem[Vanita \emph{et~al.}(2025)Vanita, Mezzadra, Tealdi, and
  Clemens]{Vanita2025}
V.~Vanita, G.~Mezzadra, C.~Tealdi and O.~Clemens, \emph{ACS Applied Energy
  Materials}, 2025, \textbf{8}, 7562--7574\relax
\mciteBstWouldAddEndPuncttrue
\mciteSetBstMidEndSepPunct{\mcitedefaultmidpunct}
{\mcitedefaultendpunct}{\mcitedefaultseppunct}\relax
\EndOfBibitem
\bibitem[Singh \emph{et~al.}(2024)Singh, K~R, and Dixit]{Singh2024}
P.~Singh, A.~M. K~R and M.~Dixit, \emph{ACS Applied Electronic Materials},
  2024, \textbf{6}, 7065--7074\relax
\mciteBstWouldAddEndPuncttrue
\mciteSetBstMidEndSepPunct{\mcitedefaultmidpunct}
{\mcitedefaultendpunct}{\mcitedefaultseppunct}\relax
\EndOfBibitem
\bibitem[Abdul~Nasir \emph{et~al.}(2025)Abdul~Nasir, Guan, Jee, Woodley, Sokol,
  Catlow, and Elena]{AbdulNasir2025}
J.~Abdul~Nasir, J.~Guan, W.~Jee, S.~M. Woodley, A.~A. Sokol, C.~R.~A. Catlow
  and A.~Elena, \emph{Physical Chemistry Chemical Physics}, 2025,
  10.1039.D5CP01882J\relax
\mciteBstWouldAddEndPuncttrue
\mciteSetBstMidEndSepPunct{\mcitedefaultmidpunct}
{\mcitedefaultendpunct}{\mcitedefaultseppunct}\relax
\EndOfBibitem
\bibitem[Kabylda \emph{et~al.}(2025)Kabylda, Mortazavi, Zhuang, and
  Tkatchenko]{Kabylda2025}
A.~Kabylda, B.~Mortazavi, X.~Zhuang and A.~Tkatchenko, \emph{Advanced
  Functional Materials}, 2025, \textbf{35}, 2417891\relax
\mciteBstWouldAddEndPuncttrue
\mciteSetBstMidEndSepPunct{\mcitedefaultmidpunct}
{\mcitedefaultendpunct}{\mcitedefaultseppunct}\relax
\EndOfBibitem
\bibitem[Demeyere \emph{et~al.}(2025)Demeyere, Ellaby, Sarwar, Thompsett, and
  Skylaris]{Demeyere2025}
T.~Demeyere, T.~Ellaby, M.~Sarwar, D.~Thompsett and C.-K. Skylaris, \emph{ACS
  Catalysis}, 2025, \textbf{15}, 5674--5682\relax
\mciteBstWouldAddEndPuncttrue
\mciteSetBstMidEndSepPunct{\mcitedefaultmidpunct}
{\mcitedefaultendpunct}{\mcitedefaultseppunct}\relax
\EndOfBibitem
\bibitem[Loew \emph{et~al.}(2025)Loew, Sun, Wang, Botti, and Marques]{Loew2025}
A.~Loew, D.~Sun, H.-C. Wang, S.~Botti and M.~A.~L. Marques, \emph{npj
  Computational Materials}, 2025, \textbf{11}, 178\relax
\mciteBstWouldAddEndPuncttrue
\mciteSetBstMidEndSepPunct{\mcitedefaultmidpunct}
{\mcitedefaultendpunct}{\mcitedefaultseppunct}\relax
\EndOfBibitem
\bibitem[Cheng \emph{et~al.}(2025)Cheng, Fu, Yu, Rha, Chotrattanapituk,
  Abernathy, Cheng, and Li]{Cheng2025}
M.~Cheng, C.-L. Fu, B.~Yu, E.~Rha, A.~Chotrattanapituk, D.~L. Abernathy,
  Y.~Cheng and M.~Li, \emph{A {Foundation} {Model} for {Non}-{Destructive}
  {Defect} {Identification} from {Vibrational} {Spectra}}, 2025,
  \url{http://arxiv.org/abs/2506.00725}, arXiv:2506.00725 [cond-mat]\relax
\mciteBstWouldAddEndPuncttrue
\mciteSetBstMidEndSepPunct{\mcitedefaultmidpunct}
{\mcitedefaultendpunct}{\mcitedefaultseppunct}\relax
\EndOfBibitem
\bibitem[Shen \emph{et~al.}(2024)Shen, Attarian, Zhang, Zhang, Asta,
  Szlufarska, and Morgan]{Shen2024}
C.~Shen, S.~Attarian, Y.~Zhang, H.~Zhang, M.~Asta, I.~Szlufarska and D.~Morgan,
  \emph{{SuperSalt}: {Equivariant} {Neural} {Network} {Force} {Fields} for
  {Multicomponent} {Molten} {Salts} {System}}, 2024,
  \url{https://arxiv.org/abs/2412.19353}, Version Number: 1\relax
\mciteBstWouldAddEndPuncttrue
\mciteSetBstMidEndSepPunct{\mcitedefaultmidpunct}
{\mcitedefaultendpunct}{\mcitedefaultseppunct}\relax
\EndOfBibitem
\bibitem[Schäfer \emph{et~al.}(2025)Schäfer, Segreto, Zills, Holm, and
  Kästner]{Schafer2025}
M.~R. Schäfer, N.~Segreto, F.~Zills, C.~Holm and J.~Kästner, \emph{Apax: {A}
  {Flexible} and {Performant} {Framework} {For} {The} {Development} of
  {Machine}-{Learned} {Interatomic} {Potentials}}, 2025,
  \url{https://arxiv.org/abs/2505.22168}, Version Number: 2\relax
\mciteBstWouldAddEndPuncttrue
\mciteSetBstMidEndSepPunct{\mcitedefaultmidpunct}
{\mcitedefaultendpunct}{\mcitedefaultseppunct}\relax
\EndOfBibitem
\bibitem[Novelli \emph{et~al.}(2025)Novelli, Meanti, Buigues, Rosasco,
  Parrinello, Pontil, and Bonati]{Novelli2025}
P.~Novelli, G.~Meanti, P.~J. Buigues, L.~Rosasco, M.~Parrinello, M.~Pontil and
  L.~Bonati, \emph{Fast and {Fourier} {Features} for {Transfer} {Learning} of
  {Interatomic} {Potentials}}, 2025, \url{https://arxiv.org/abs/2505.05652},
  Version Number: 1\relax
\mciteBstWouldAddEndPuncttrue
\mciteSetBstMidEndSepPunct{\mcitedefaultmidpunct}
{\mcitedefaultendpunct}{\mcitedefaultseppunct}\relax
\EndOfBibitem
\bibitem[Hörmann \emph{et~al.}(2025)Hörmann, Stark, and Maurer]{Hormann2025}
L.~Hörmann, W.~G. Stark and R.~J. Maurer, \emph{npj Computational Materials},
  2025, \textbf{11}, 196\relax
\mciteBstWouldAddEndPuncttrue
\mciteSetBstMidEndSepPunct{\mcitedefaultmidpunct}
{\mcitedefaultendpunct}{\mcitedefaultseppunct}\relax
\EndOfBibitem
\bibitem[Schwalbe-Koda \emph{et~al.}(2025)Schwalbe-Koda, Govindarajan, and
  Varley]{Schwalbe-Koda2025}
D.~Schwalbe-Koda, N.~Govindarajan and J.~B. Varley, \emph{Digital Discovery},
  2025, \textbf{4}, 234--251\relax
\mciteBstWouldAddEndPuncttrue
\mciteSetBstMidEndSepPunct{\mcitedefaultmidpunct}
{\mcitedefaultendpunct}{\mcitedefaultseppunct}\relax
\EndOfBibitem
\bibitem[Gupta \emph{et~al.}(2025)Gupta, Rajan, Fako, Gonçalves, Müller,
  Varghese, Schäfer, and De]{Gupta2025}
S.~Gupta, A.~Rajan, E.~Fako, T.~J.~F. Gonçalves, I.~B. Müller, J.~J.
  Varghese, A.~Schäfer and S.~De, \emph{The Journal of Physical Chemistry C},
  2025, \textbf{129}, 3022--3033\relax
\mciteBstWouldAddEndPuncttrue
\mciteSetBstMidEndSepPunct{\mcitedefaultmidpunct}
{\mcitedefaultendpunct}{\mcitedefaultseppunct}\relax
\EndOfBibitem
\bibitem[Fako and De(2025)]{Fako2025}
E.~Fako and S.~De, \emph{Simple {Heuristics} for {Advanced} {Sampling} of
  {Reactive} {Species} on {Surfaces}}, 2025,
  \url{https://chemrxiv.org/engage/chemrxiv/article-details/67f8132181d2151a02dfba51}\relax
\mciteBstWouldAddEndPuncttrue
\mciteSetBstMidEndSepPunct{\mcitedefaultmidpunct}
{\mcitedefaultendpunct}{\mcitedefaultseppunct}\relax
\EndOfBibitem
\bibitem[Tian \emph{et~al.}(2025)Tian, Tosello~Gardini, Raucci, Xiao, Zhuo, and
  Parrinello]{Tian2025}
X.~Tian, A.~Tosello~Gardini, U.~Raucci, H.~Xiao, Y.~Zhuo and M.~Parrinello,
  \emph{Electrochemical {Potential}-{Driven} {Water} {Dynamics} {Control} {CO2}
  {Electroreduction} at the {Ag}/{H2O} {Interface}}, 2025,
  \url{https://chemrxiv.org/engage/chemrxiv/article-details/68331a69c1cb1ecda051890b}\relax
\mciteBstWouldAddEndPuncttrue
\mciteSetBstMidEndSepPunct{\mcitedefaultmidpunct}
{\mcitedefaultendpunct}{\mcitedefaultseppunct}\relax
\EndOfBibitem
\bibitem[Pitfield \emph{et~al.}(2025)Pitfield, Christiansen, and
  Hammer]{Pitfield2025}
J.~Pitfield, M.-P.~V. Christiansen and B.~Hammer, \emph{Active
  \${\textbackslash}{Delta}\$-learning with universal potentials for global
  structure optimization}, 2025, \url{https://arxiv.org/abs/2507.18485},
  Version Number: 1\relax
\mciteBstWouldAddEndPuncttrue
\mciteSetBstMidEndSepPunct{\mcitedefaultmidpunct}
{\mcitedefaultendpunct}{\mcitedefaultseppunct}\relax
\EndOfBibitem
\bibitem[Christiansen and Hammer(2025)]{Christiansen2025}
M.-P.~V. Christiansen and B.~Hammer, \emph{The Journal of Chemical Physics},
  2025, \textbf{162}, 184701\relax
\mciteBstWouldAddEndPuncttrue
\mciteSetBstMidEndSepPunct{\mcitedefaultmidpunct}
{\mcitedefaultendpunct}{\mcitedefaultseppunct}\relax
\EndOfBibitem
\bibitem[Soyemi \emph{et~al.}(2025)Soyemi, Baral, and Szilvasi]{Soyemi2025}
A.~Soyemi, K.~Baral and T.~Szilvasi, \emph{A {Simple} {Iterative} {Approach}
  for {Constant} {Chemical} {Potential} {Simulations} at {Interfaces}}, 2025,
  \url{https://arxiv.org/abs/2506.01050}, Version Number: 1\relax
\mciteBstWouldAddEndPuncttrue
\mciteSetBstMidEndSepPunct{\mcitedefaultmidpunct}
{\mcitedefaultendpunct}{\mcitedefaultseppunct}\relax
\EndOfBibitem
\bibitem[Cvitkovich \emph{et~al.}(2024)Cvitkovich, Fehringer, Wilhelmer,
  Milardovich, Waldhör, and Grasser]{Cvitkovich2024}
L.~Cvitkovich, F.~Fehringer, C.~Wilhelmer, D.~Milardovich, D.~Waldhör and
  T.~Grasser, \emph{The Journal of Chemical Physics}, 2024, \textbf{161},
  144706\relax
\mciteBstWouldAddEndPuncttrue
\mciteSetBstMidEndSepPunct{\mcitedefaultmidpunct}
{\mcitedefaultendpunct}{\mcitedefaultseppunct}\relax
\EndOfBibitem
\bibitem[Deng \emph{et~al.}(2025)Deng, Choi, Zhong, Riebesell, Anand, Li, Jun,
  Persson, and Ceder]{Deng2025}
B.~Deng, Y.~Choi, P.~Zhong, J.~Riebesell, S.~Anand, Z.~Li, K.~Jun, K.~A.
  Persson and G.~Ceder, \emph{npj Computational Materials}, 2025, \textbf{11},
  9\relax
\mciteBstWouldAddEndPuncttrue
\mciteSetBstMidEndSepPunct{\mcitedefaultmidpunct}
{\mcitedefaultendpunct}{\mcitedefaultseppunct}\relax
\EndOfBibitem
\bibitem[Rensmeyer \emph{et~al.}(2025)Rensmeyer, Kramer, and
  Niggemann]{Rensmeyer2025}
T.~Rensmeyer, D.~Kramer and O.~Niggemann, \emph{On-the-{Fly} {Fine}-{Tuning} of
  {Foundational} {Neural} {Network} {Potentials}: {A} {Bayesian} {Neural}
  {Network} {Approach}}, 2025, \url{http://arxiv.org/abs/2507.13805},
  arXiv:2507.13805 [cs]\relax
\mciteBstWouldAddEndPuncttrue
\mciteSetBstMidEndSepPunct{\mcitedefaultmidpunct}
{\mcitedefaultendpunct}{\mcitedefaultseppunct}\relax
\EndOfBibitem
\bibitem[Bertani and Pedone(2025)]{Bertani2025}
M.~Bertani and A.~Pedone, \emph{The Journal of Physical Chemistry C}, 2025,
  \textbf{129}, 12697--12709\relax
\mciteBstWouldAddEndPuncttrue
\mciteSetBstMidEndSepPunct{\mcitedefaultmidpunct}
{\mcitedefaultendpunct}{\mcitedefaultseppunct}\relax
\EndOfBibitem
\bibitem[Hänseroth and Dreßler(2025)]{Hanseroth2025}
J.~Hänseroth and C.~Dreßler, \emph{Optimizing {Machine} {Learning}
  {Potentials} for {Hydroxide} {Transport}: {Surprising} {Efficiency} of
  {Single}-{Concentration} {Training}}, 2025,
  \url{https://arxiv.org/abs/2505.07580}, Version Number: 1\relax
\mciteBstWouldAddEndPuncttrue
\mciteSetBstMidEndSepPunct{\mcitedefaultmidpunct}
{\mcitedefaultendpunct}{\mcitedefaultseppunct}\relax
\EndOfBibitem
\bibitem[Kaur \emph{et~al.}(2025)Kaur, Pia, Batatia, Advincula, Shi, Lan,
  Csányi, Michaelides, and Kapil]{Kaur2025}
H.~Kaur, F.~D. Pia, I.~Batatia, X.~R. Advincula, B.~X. Shi, J.~Lan, G.~Csányi,
  A.~Michaelides and V.~Kapil, \emph{Faraday Discussions}, 2025, \textbf{256},
  120--138\relax
\mciteBstWouldAddEndPuncttrue
\mciteSetBstMidEndSepPunct{\mcitedefaultmidpunct}
{\mcitedefaultendpunct}{\mcitedefaultseppunct}\relax
\EndOfBibitem
\bibitem[Elena \emph{et~al.}(2025)Elena, Kamath, Jaffrelot~Inizan, Rosen,
  Zanca, and Persson]{Elena2025}
A.~M. Elena, P.~D. Kamath, T.~Jaffrelot~Inizan, A.~S. Rosen, F.~Zanca and K.~A.
  Persson, \emph{npj Computational Materials}, 2025, \textbf{11}, 125\relax
\mciteBstWouldAddEndPuncttrue
\mciteSetBstMidEndSepPunct{\mcitedefaultmidpunct}
{\mcitedefaultendpunct}{\mcitedefaultseppunct}\relax
\EndOfBibitem
\bibitem[Lim \emph{et~al.}(2025)Lim, Park, Walsh, and Kim]{Lim2025}
Y.~Lim, H.~Park, A.~Walsh and J.~Kim, \emph{Matter}, 2025, \textbf{8},
  102203\relax
\mciteBstWouldAddEndPuncttrue
\mciteSetBstMidEndSepPunct{\mcitedefaultmidpunct}
{\mcitedefaultendpunct}{\mcitedefaultseppunct}\relax
\EndOfBibitem
\bibitem[Steffen(2025)]{Steffen2025}
J.~Steffen, \emph{The Journal of Physical Chemistry C}, 2025, \textbf{129},
  13513--13531\relax
\mciteBstWouldAddEndPuncttrue
\mciteSetBstMidEndSepPunct{\mcitedefaultmidpunct}
{\mcitedefaultendpunct}{\mcitedefaultseppunct}\relax
\EndOfBibitem
\bibitem[Focassio \emph{et~al.}(2025)Focassio, M.~Freitas, and
  Schleder]{Focassio2025}
B.~Focassio, L.~P. M.~Freitas and G.~R. Schleder, \emph{ACS Applied Materials
  \& Interfaces}, 2025, \textbf{17}, 13111--13121\relax
\mciteBstWouldAddEndPuncttrue
\mciteSetBstMidEndSepPunct{\mcitedefaultmidpunct}
{\mcitedefaultendpunct}{\mcitedefaultseppunct}\relax
\EndOfBibitem
\bibitem[Allam \emph{et~al.}(2024)Allam, Maghsoodi, Jang, and Snow]{Allam2024}
O.~Allam, M.~Maghsoodi, S.~S. Jang and S.~D. Snow, \emph{ACS Applied Materials
  \&amp; Interfaces}, 2024, \textbf{16}, 36215--36223\relax
\mciteBstWouldAddEndPuncttrue
\mciteSetBstMidEndSepPunct{\mcitedefaultmidpunct}
{\mcitedefaultendpunct}{\mcitedefaultseppunct}\relax
\EndOfBibitem
\bibitem[Boulangeot \emph{et~al.}(2024)Boulangeot, Brix, Sur, and
  Gaudry]{Boulangeot2024}
N.~Boulangeot, F.~Brix, F.~Sur and E.~Gaudry, \emph{Journal of Chemical Theory
  and Computation}, 2024,  acs.jctc.4c00367\relax
\mciteBstWouldAddEndPuncttrue
\mciteSetBstMidEndSepPunct{\mcitedefaultmidpunct}
{\mcitedefaultendpunct}{\mcitedefaultseppunct}\relax
\EndOfBibitem
\bibitem[Du \emph{et~al.}(2025)Du, Liu, Peng, Chun, Hoffman, Yildiz, Li,
  Bazant, and Gómez-Bombarelli]{Du2025}
X.~Du, M.~Liu, J.~Peng, H.~Chun, A.~Hoffman, B.~Yildiz, L.~Li, M.~Z. Bazant and
  R.~Gómez-Bombarelli, \emph{ACS Central Science}, 2025,
  acscentsci.5c00547\relax
\mciteBstWouldAddEndPuncttrue
\mciteSetBstMidEndSepPunct{\mcitedefaultmidpunct}
{\mcitedefaultendpunct}{\mcitedefaultseppunct}\relax
\EndOfBibitem
\bibitem[Pieck(2021)]{Pieck2021a}
F.~Pieck, \emph{{NOMAD} dataset: {Bonding} and reactivity of an
  alkyne-functionalized cyclooctyne on {Si}(001) with quantitative electronic
  structure analysis}, 2021,
  \url{https://nomad-lab.eu/prod/v1/gui/dataset/doi/10.17172/NOMAD/2021.09.28-2}\relax
\mciteBstWouldAddEndPuncttrue
\mciteSetBstMidEndSepPunct{\mcitedefaultmidpunct}
{\mcitedefaultendpunct}{\mcitedefaultseppunct}\relax
\EndOfBibitem
\bibitem[Kresse and Hafner(1993)]{Kresse1993}
G.~Kresse and J.~Hafner, \emph{Physical Review B}, 1993, \textbf{47},
  558--561\relax
\mciteBstWouldAddEndPuncttrue
\mciteSetBstMidEndSepPunct{\mcitedefaultmidpunct}
{\mcitedefaultendpunct}{\mcitedefaultseppunct}\relax
\EndOfBibitem
\bibitem[Kresse and Hafner(1994)]{Kresse1994}
G.~Kresse and J.~Hafner, \emph{Physical Review B}, 1994, \textbf{49},
  14251--14269\relax
\mciteBstWouldAddEndPuncttrue
\mciteSetBstMidEndSepPunct{\mcitedefaultmidpunct}
{\mcitedefaultendpunct}{\mcitedefaultseppunct}\relax
\EndOfBibitem
\bibitem[Kresse and Furthmüller(1996)]{Kresse1996}
G.~Kresse and J.~Furthmüller, \emph{Physical Review B}, 1996, \textbf{54},
  11169--11186\relax
\mciteBstWouldAddEndPuncttrue
\mciteSetBstMidEndSepPunct{\mcitedefaultmidpunct}
{\mcitedefaultendpunct}{\mcitedefaultseppunct}\relax
\EndOfBibitem
\bibitem[Kresse and Furthmüller(1996)]{Kresse1996a}
G.~Kresse and J.~Furthmüller, \emph{Computational Materials Science}, 1996,
  \textbf{6}, 15--50\relax
\mciteBstWouldAddEndPuncttrue
\mciteSetBstMidEndSepPunct{\mcitedefaultmidpunct}
{\mcitedefaultendpunct}{\mcitedefaultseppunct}\relax
\EndOfBibitem
\bibitem[Kresse and Joubert(1999)]{Kresse1999}
G.~Kresse and D.~Joubert, \emph{Physical Review B}, 1999, \textbf{59},
  1758--1775\relax
\mciteBstWouldAddEndPuncttrue
\mciteSetBstMidEndSepPunct{\mcitedefaultmidpunct}
{\mcitedefaultendpunct}{\mcitedefaultseppunct}\relax
\EndOfBibitem
\bibitem[Perdew \emph{et~al.}(1996)Perdew, Burke, and Ernzerhof]{Perdew1996}
J.~P. Perdew, K.~Burke and M.~Ernzerhof, \emph{Physical Review Letters}, 1996,
  \textbf{77}, 3865--3868\relax
\mciteBstWouldAddEndPuncttrue
\mciteSetBstMidEndSepPunct{\mcitedefaultmidpunct}
{\mcitedefaultendpunct}{\mcitedefaultseppunct}\relax
\EndOfBibitem
\bibitem[Perdew \emph{et~al.}(1997)Perdew, Burke, and Ernzerhof]{Perdew1997}
J.~P. Perdew, K.~Burke and M.~Ernzerhof, \emph{Physical Review Letters}, 1997,
  \textbf{78}, 1396--1396\relax
\mciteBstWouldAddEndPuncttrue
\mciteSetBstMidEndSepPunct{\mcitedefaultmidpunct}
{\mcitedefaultendpunct}{\mcitedefaultseppunct}\relax
\EndOfBibitem
\bibitem[Grimme \emph{et~al.}(2010)Grimme, Antony, Ehrlich, and
  Krieg]{Grimme2010}
S.~Grimme, J.~Antony, S.~Ehrlich and H.~Krieg, \emph{The Journal of Chemical
  Physics}, 2010, \textbf{132}, 154104\relax
\mciteBstWouldAddEndPuncttrue
\mciteSetBstMidEndSepPunct{\mcitedefaultmidpunct}
{\mcitedefaultendpunct}{\mcitedefaultseppunct}\relax
\EndOfBibitem
\bibitem[Grimme \emph{et~al.}(2011)Grimme, Ehrlich, and Goerigk]{Grimme2011}
S.~Grimme, S.~Ehrlich and L.~Goerigk, \emph{Journal of Computational
  Chemistry}, 2011, \textbf{32}, 1456--1465\relax
\mciteBstWouldAddEndPuncttrue
\mciteSetBstMidEndSepPunct{\mcitedefaultmidpunct}
{\mcitedefaultendpunct}{\mcitedefaultseppunct}\relax
\EndOfBibitem
\bibitem[Nosé(1984)]{Nose1984}
S.~Nosé, \emph{The Journal of Chemical Physics}, 1984, \textbf{81},
  511--519\relax
\mciteBstWouldAddEndPuncttrue
\mciteSetBstMidEndSepPunct{\mcitedefaultmidpunct}
{\mcitedefaultendpunct}{\mcitedefaultseppunct}\relax
\EndOfBibitem
\bibitem[Hoover(1985)]{Hoover1985}
W.~G. Hoover, \emph{Physical Review A}, 1985, \textbf{31}, 1695--1697\relax
\mciteBstWouldAddEndPuncttrue
\mciteSetBstMidEndSepPunct{\mcitedefaultmidpunct}
{\mcitedefaultendpunct}{\mcitedefaultseppunct}\relax
\EndOfBibitem
\bibitem[Nosé(1991)]{Nose1991}
S.~Nosé, \emph{Progress of Theoretical Physics Supplement}, 1991,
  \textbf{103}, 1--46\relax
\mciteBstWouldAddEndPuncttrue
\mciteSetBstMidEndSepPunct{\mcitedefaultmidpunct}
{\mcitedefaultendpunct}{\mcitedefaultseppunct}\relax
\EndOfBibitem
\bibitem[Tiefenbacher \emph{et~al.}(2025)Tiefenbacher, Bachmair, Chen,
  Westermayr, Marquetand, Dietschreit, and González]{Tiefenbacher2025}
M.~X. Tiefenbacher, B.~Bachmair, C.~G. Chen, J.~Westermayr, P.~Marquetand,
  J.~C.~B. Dietschreit and L.~González, \emph{Digital Discovery}, 2025,
  \textbf{4}, 1478--1491\relax
\mciteBstWouldAddEndPuncttrue
\mciteSetBstMidEndSepPunct{\mcitedefaultmidpunct}
{\mcitedefaultendpunct}{\mcitedefaultseppunct}\relax
\EndOfBibitem
\bibitem[Smith \emph{et~al.}(2018)Smith, Nebgen, Lubbers, Isayev, and
  Roitberg]{Smith2018}
J.~S. Smith, B.~Nebgen, N.~Lubbers, O.~Isayev and A.~E. Roitberg, \emph{The
  Journal of Chemical Physics}, 2018, \textbf{148}, 241733\relax
\mciteBstWouldAddEndPuncttrue
\mciteSetBstMidEndSepPunct{\mcitedefaultmidpunct}
{\mcitedefaultendpunct}{\mcitedefaultseppunct}\relax
\EndOfBibitem
\bibitem[Larsen \emph{et~al.}(2017)Larsen, Mortensen, Blomqvist, Castelli,
  Christensen, Dułak, Friis, Groves, Hammer, Hargus, Hermes, Jennings, Jensen,
  Kermode, Kitchin, Kolsbjerg, Kubal, Kaasbjerg, Lysgaard, Maronsson, Maxson,
  Olsen, Pastewka, Peterson, Rostgaard, Schiøtz, Schütt, Strange, Thygesen,
  Vegge, Vilhelmsen, Walter, Zeng, and Jacobsen]{Larsen2017}
A.~H. Larsen, J.~J. Mortensen, J.~Blomqvist, I.~E. Castelli, R.~Christensen,
  M.~Dułak, J.~Friis, M.~N. Groves, B.~Hammer, C.~Hargus, E.~D. Hermes, P.~C.
  Jennings, P.~B. Jensen, J.~Kermode, J.~R. Kitchin, E.~L. Kolsbjerg, J.~Kubal,
  K.~Kaasbjerg, S.~Lysgaard, J.~B. Maronsson, T.~Maxson, T.~Olsen, L.~Pastewka,
  A.~Peterson, C.~Rostgaard, J.~Schiøtz, O.~Schütt, M.~Strange, K.~S.
  Thygesen, T.~Vegge, L.~Vilhelmsen, M.~Walter, Z.~Zeng and K.~W. Jacobsen,
  \emph{Journal of Physics: Condensed Matter}, 2017, \textbf{29},
  273002--273002\relax
\mciteBstWouldAddEndPuncttrue
\mciteSetBstMidEndSepPunct{\mcitedefaultmidpunct}
{\mcitedefaultendpunct}{\mcitedefaultseppunct}\relax
\EndOfBibitem
\bibitem[Bigi \emph{et~al.}(2024)Bigi, Langer, and Ceriotti]{Bigi2024}
F.~Bigi, M.~Langer and M.~Ceriotti, \emph{The dark side of the forces:
  assessing non-conservative force models for atomistic machine learning},
  2024, \url{https://arxiv.org/abs/2412.11569}, Version Number: 5\relax
\mciteBstWouldAddEndPuncttrue
\mciteSetBstMidEndSepPunct{\mcitedefaultmidpunct}
{\mcitedefaultendpunct}{\mcitedefaultseppunct}\relax
\EndOfBibitem
\bibitem[Deng \emph{et~al.}(2023)Deng, Zhong, Jun, Riebesell, Han, Bartel, and
  Ceder]{Deng2023a}
B.~Deng, P.~Zhong, K.~Jun, J.~Riebesell, K.~Han, C.~J. Bartel and G.~Ceder,
  \emph{Nature Machine Intelligence}, 2023, \textbf{5}, 1031--1041\relax
\mciteBstWouldAddEndPuncttrue
\mciteSetBstMidEndSepPunct{\mcitedefaultmidpunct}
{\mcitedefaultendpunct}{\mcitedefaultseppunct}\relax
\EndOfBibitem
\bibitem[MAC()]{MACE_manual}
\emph{Fine-tuning {Foundation} {Models} — mace 0.3.13 documentation},
  \url{https://mace-docs.readthedocs.io/en/latest/guide/finetuning.html#naive-fine-tuning}\relax
\mciteBstWouldAddEndPuncttrue
\mciteSetBstMidEndSepPunct{\mcitedefaultmidpunct}
{\mcitedefaultendpunct}{\mcitedefaultseppunct}\relax
\EndOfBibitem
\bibitem[Melix \emph{et~al.}(2019)Melix, Paesani, and Heine]{Melix2019e}
P.~Melix, F.~Paesani and T.~Heine, \emph{Advanced Theory and Simulations},
  2019, \textbf{2}, 1900098--1900098\relax
\mciteBstWouldAddEndPuncttrue
\mciteSetBstMidEndSepPunct{\mcitedefaultmidpunct}
{\mcitedefaultendpunct}{\mcitedefaultseppunct}\relax
\EndOfBibitem
\bibitem[Community(2018)]{Blender}
B.~O. Community, \emph{Blender - a {3D} modelling and rendering package}, 2018,
  \url{http://www.blender.org}\relax
\mciteBstWouldAddEndPuncttrue
\mciteSetBstMidEndSepPunct{\mcitedefaultmidpunct}
{\mcitedefaultendpunct}{\mcitedefaultseppunct}\relax
\EndOfBibitem
\bibitem[Weiske \emph{et~al.}(2024)Weiske, Thiemann, and
  Melix]{Weiske_Blender2024}
H.~Weiske, F.~Thiemann and P.~Melix, \emph{Blender {Import} {ASE}}, 2024,
  \url{https://zenodo.org/doi/10.5281/zenodo.10776697}\relax
\mciteBstWouldAddEndPuncttrue
\mciteSetBstMidEndSepPunct{\mcitedefaultmidpunct}
{\mcitedefaultendpunct}{\mcitedefaultseppunct}\relax
\EndOfBibitem
\bibitem[Harris \emph{et~al.}(2020)Harris, Millman, Van Der~Walt, Gommers,
  Virtanen, Cournapeau, Wieser, Taylor, Berg, Smith, Kern, Picus, Hoyer,
  Van~Kerkwijk, Brett, Haldane, Del~Río, Wiebe, Peterson, Gérard-Marchant,
  Sheppard, Reddy, Weckesser, Abbasi, Gohlke, and Oliphant]{Harris2020}
C.~R. Harris, K.~J. Millman, S.~J. Van Der~Walt, R.~Gommers, P.~Virtanen,
  D.~Cournapeau, E.~Wieser, J.~Taylor, S.~Berg, N.~J. Smith, R.~Kern, M.~Picus,
  S.~Hoyer, M.~H. Van~Kerkwijk, M.~Brett, A.~Haldane, J.~F. Del~Río, M.~Wiebe,
  P.~Peterson, P.~Gérard-Marchant, K.~Sheppard, T.~Reddy, W.~Weckesser,
  H.~Abbasi, C.~Gohlke and T.~E. Oliphant, \emph{Nature}, 2020, \textbf{585},
  357--362\relax
\mciteBstWouldAddEndPuncttrue
\mciteSetBstMidEndSepPunct{\mcitedefaultmidpunct}
{\mcitedefaultendpunct}{\mcitedefaultseppunct}\relax
\EndOfBibitem
\bibitem[Pecher and Tonner(2018)]{pecher_bond_2018}
L.~Pecher and R.~Tonner, \emph{Inorganics}, 2018, \textbf{6}, 17\relax
\mciteBstWouldAddEndPuncttrue
\mciteSetBstMidEndSepPunct{\mcitedefaultmidpunct}
{\mcitedefaultendpunct}{\mcitedefaultseppunct}\relax
\EndOfBibitem
\bibitem[Weiske \emph{et~al.}(2025)Weiske, Barett, Tonner-Zech, Melix, and
  Westermayr]{Weiske2025_data}
H.~Weiske, R.~Barett, R.~Tonner-Zech, P.~Melix and J.~Westermayr,
  \emph{Dataset: {Statistically} {Relevant} {Adsorption} {Dynamics} of {ECCO}
  on {Si}(001) using a {Fine}-{Tuned} {Foundational} {Machine}-{Learning}
  {Model}}, 2025, \url{https://doi.org/10.5281/zenodo.16836065}\relax
\mciteBstWouldAddEndPuncttrue
\mciteSetBstMidEndSepPunct{\mcitedefaultmidpunct}
{\mcitedefaultendpunct}{\mcitedefaultseppunct}\relax
\EndOfBibitem
\end{mcitethebibliography}
\end{document}


\thispagestyle{plain}


 \noindent\LARGE{\textbf {\\ Supplementary Information for "Statistics makes a difference: Machine learning adsorption dynamics of functionalized cyclooctine on Si(001) at DFT accuracy"\\}
} 

 \noindent\large{Hendrik Weiske,\textit{$^{a}$} Rhyan Barrett,\textit{$^{a}$} Ralf Tonner-Zech,\textit{$^{a}$} Patrick Melix,\textit{$^{a}$} and Julia Westermayr$^{\ast}$\textit{$^{a,b}$ \\}} \\
\textit{$^{a}$~Wilhelm Ostwald Institute, Leipzig University, Leipzig, Germany\\ $^{b}$~Center for Scalable Data Analytics and Artificial Intelligence Dresden/Leipzig, Leipzig, Germany\\ E-mail: julia.westermayr@uni-leipzig.de\\~\\ }
\noindent\normalsize{
\tableofcontents

\renewcommand{\thefigure}{S\arabic{figure}} 
\renewcommand{\thetable}{S\arabic{table}} 
\renewcommand{\thepage}{S\arabic{page}} 
\renewcommand{\thesection}{S\arabic{section}}
\newcommand*\mycommand[1]{\texttt{\emph{#1}}}
%


\clearpage
\section{Full Statistics of the Models}
\begin{figure*}[h]
   \centering
   \includegraphics[width=1\textwidth]{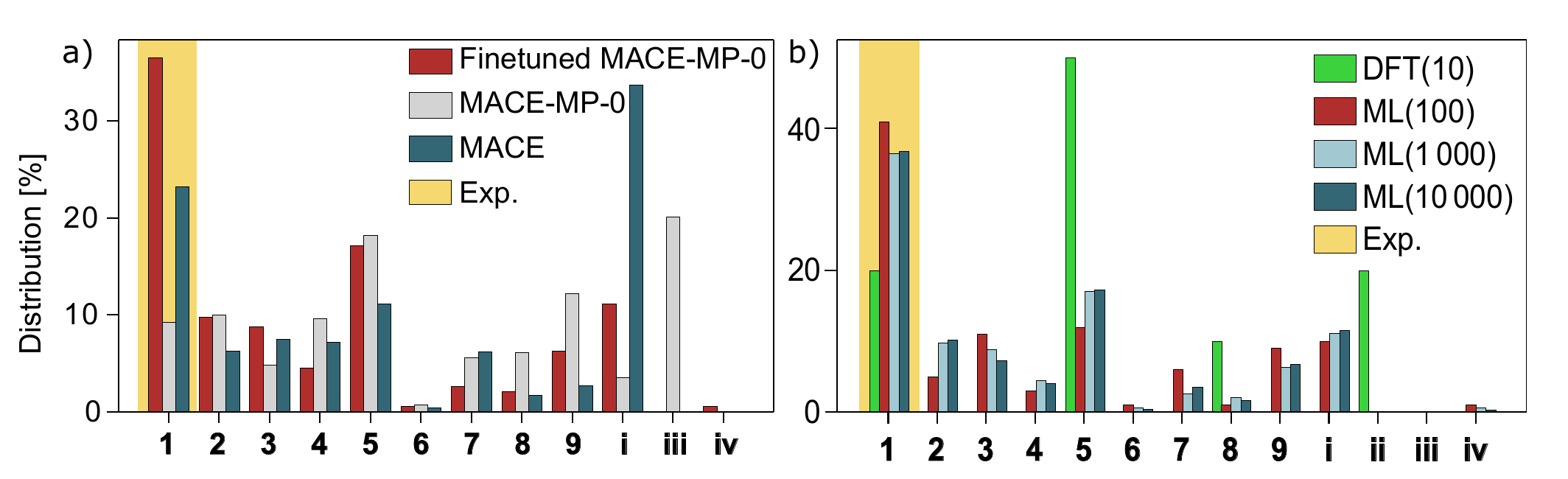}
   \caption{a) Distribution of the binding modes at the end of each MD simulation for the discussed ML models and the AIMD reference. The binding mode distribution of the fine-tuned model is given in red, the MACE-MP-0 model in grey, and the MACE model trained from scratch in dark blue. The experimentally observed binding mode is highlighted in yellow. b) Convergence of the distribution of adsorption sites for 10 (DFT) 100 (ML) 1\,000 (ML) and 10\,000 (ML) trajectories. The ML-driven trajectories are obtained using fine-tuned MACE-MP-0. Experimental observations are marked in yellow.}
   \label{Sfig:barplots}
\end{figure*}

\begin{figure}[h]
    \centering
    \includegraphics[width=0.75\linewidth]{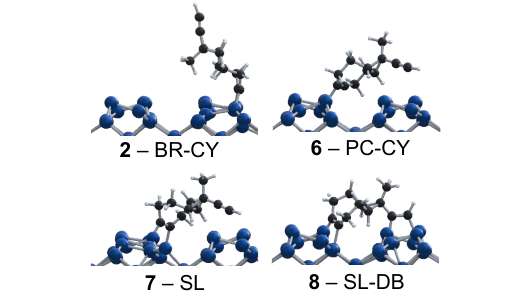}
    \caption{Binding modes detected at the end of MDs conducted using the fine-tuned MACE-MP-0 model, which are not observed in the reference AIMDs. Si atoms blue, C black and H white.}
    \label{fig:ML_not_DFT}
\end{figure}

Several binding modes not observed as final states in AIMD trajectories are sampled in our machine learning-driven MDs (see Figure~\ref{fig:ML_not_DFT}). Closer inspection of the AIMD trajectories, however, reveals these structures as transient intermediates. For instance, the prevalent doubly bound structure (\textbf{5}) is typically preceded by singly bound states. Binding modes BR-CY (\textbf{2}) and SL (\textbf{7}), as well as the PC (\textbf{6}) states, are found along AIMD trajectories as intermediates. The doubly bound sublayer structure SL-DB (\textbf{8}) and double-PC (\textbf{9}) states are the only modes detected in ML-MDs which do not appear in the DFT data.

\clearpage
\section{ML Model Comparison}
\begin{table*}[h]
    \centering
     \caption{Mean absolute error of the energies and forces for different models. \textsuperscript{a}Fine-Tuned model with the production/half/quarter of the production data. \textsuperscript{b}Foundational MACE-MP-0 model without fine tuning. \textsuperscript{c}From-scratch training without foundational model.}
    \includegraphics{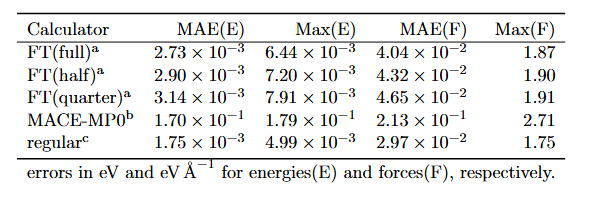}
    \label{Stab:MAE different models}
\end{table*}
\begin{table*}[h]
    \centering
    \caption{Proportions test of of 1,000 vs 10,000 trajectories.}
    \includegraphics{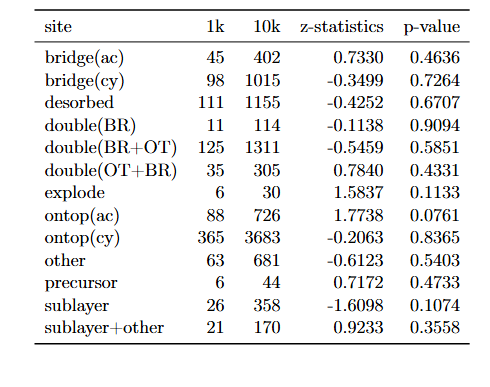}
    \label{stab:proportions}
\end{table*}

\begin{figure}[h]
    \centering
    \includegraphics[width=\columnwidth]{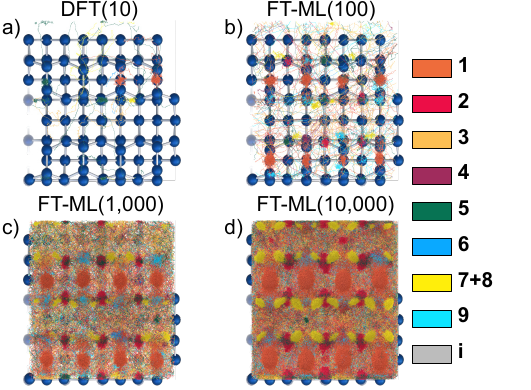}
    \caption{Median position of the cyclooctyne triple bond over all simulation runs for the 10 DFT (a), 100 FT (b) 1000 FT (c) and 10,000 FT (d) MDs. The bins are created with a size of $0.0015\,$\AA$^2$. Coloring according to the final adsorption mode of trajectory.}
    \label{Sfig:ECCO_sampling_sites}
\end{figure}

\begin{figure}[!htb]
    \centering
    \includegraphics[width=1\columnwidth]{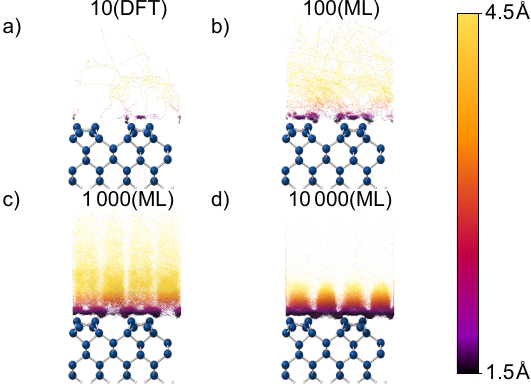}
    \caption{Representation of the sampling (side-view). The median positions of the cyclooctyne triple bond over all MD runs for the 10 DFT (a), 100 ML (b) 1000 ML (c) and 10,000 (d). The bins are created with a size of $0.0015\,$\AA$^2$ and colored using their z-value. FT refers to fine-tuned models.}
    \label{Sfig:SI_sampling_surface_side}
\end{figure}

\begin{figure}[h]
    \centering
    \includegraphics[width=\columnwidth]{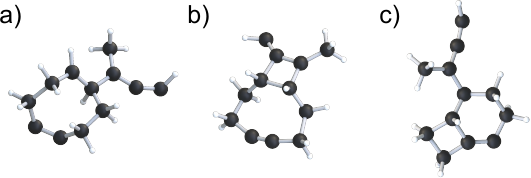}
    \caption{Examplary broken molecule structures, occurring using MACE-MP-0 without fine-tuning. Shown are a broken cyclopropanyl-structure (a), formation of a cyclobutanyl-ring at the acetylene group (b) and formation of a cyclobutanyl-ring at the cyclooctyne tripple bond (c)}
    \label{Sfig:broken molecules}
\end{figure}


\section{PCA analysis resolved by energy}
\begin{figure}[h]
    \centering
    \includegraphics[width=0.49\linewidth]{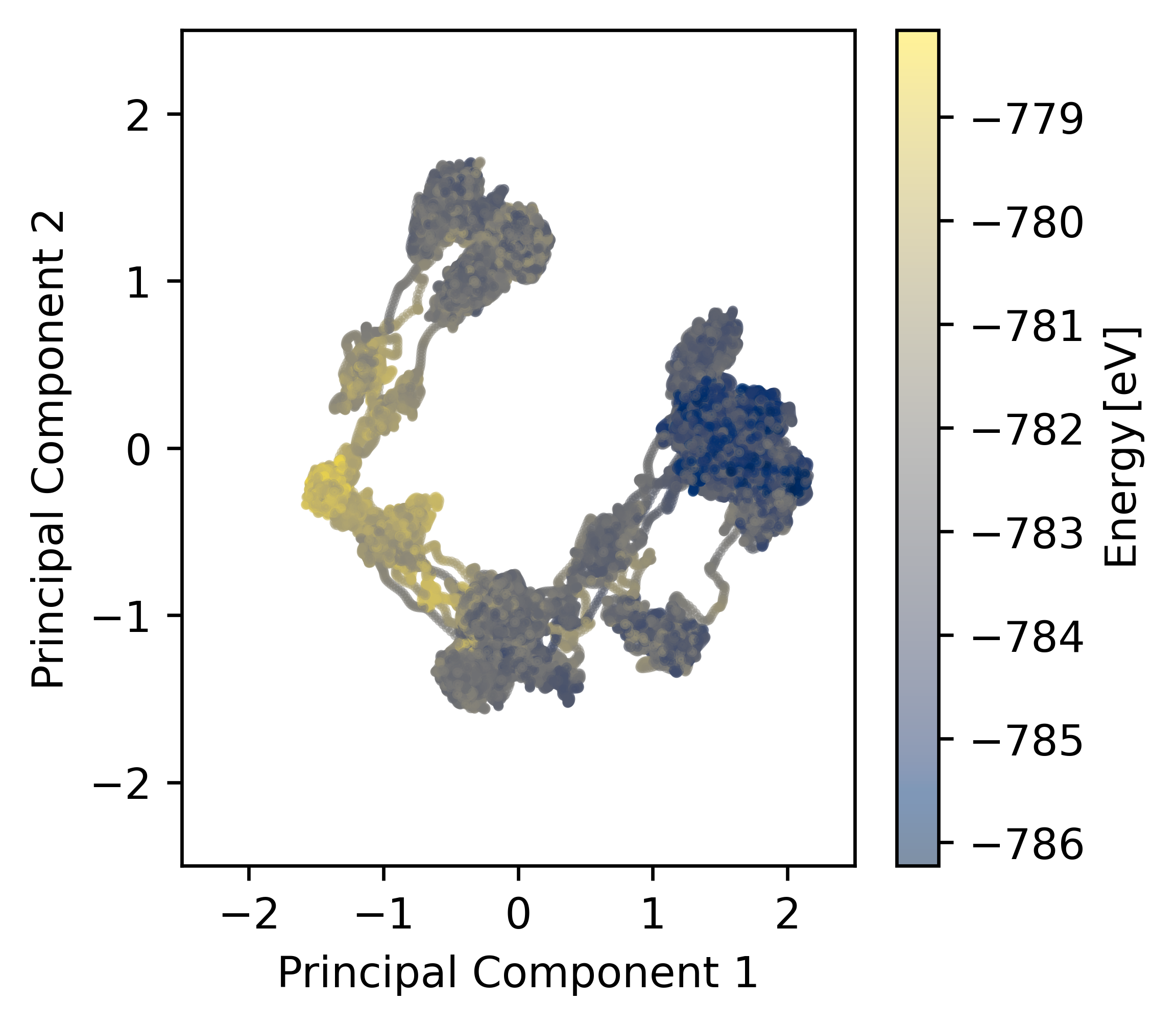}
    \includegraphics[width=0.49\linewidth]{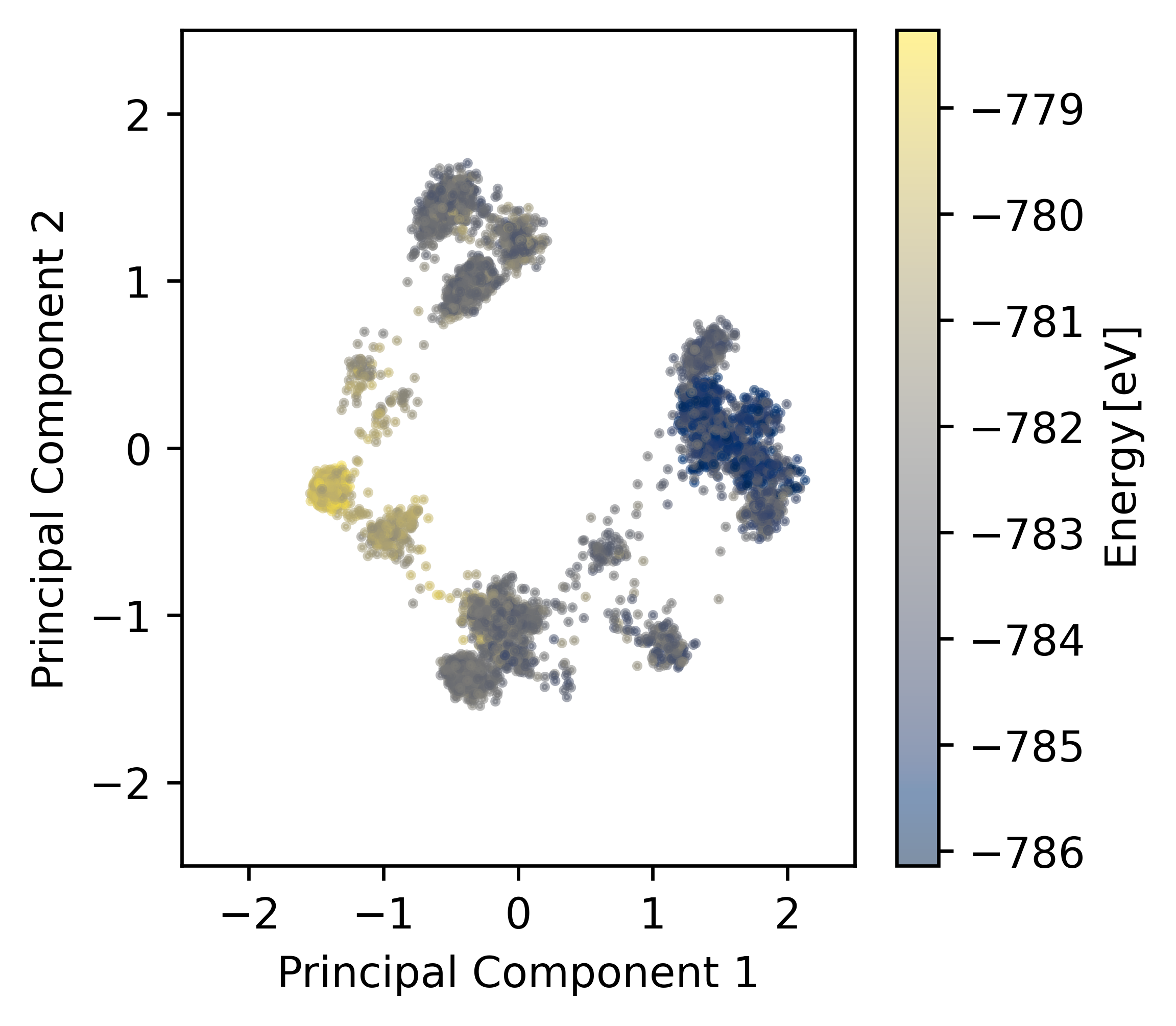}
    \caption{PCA analysis of full DFT (left) and production DFT (right) dataset using MACE descriptors.}
    \label{Sfig:PCA_DFT}
\end{figure}
\begin{figure}[h]
    \centering
    \includegraphics[width=0.49\linewidth]{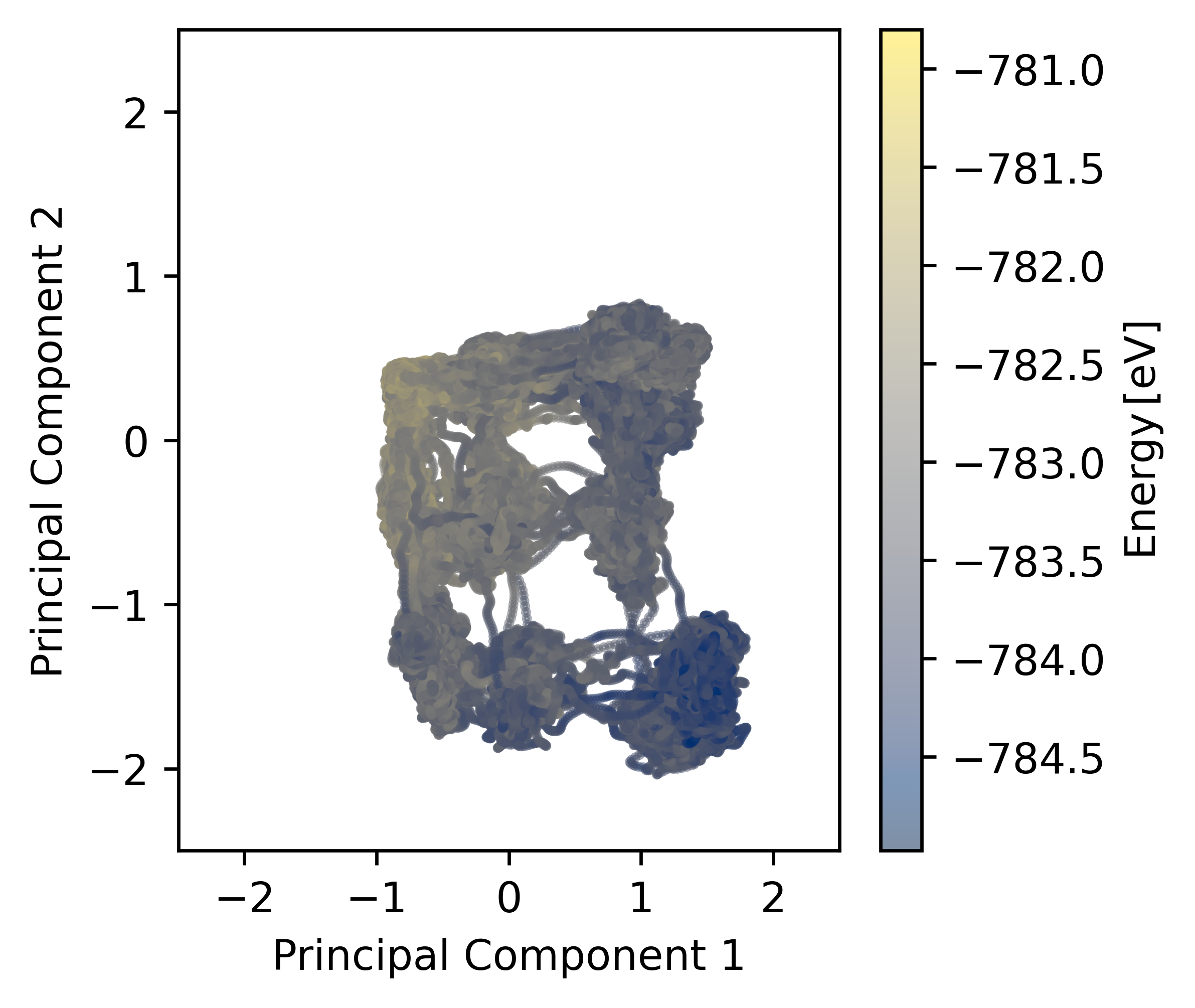}
    \includegraphics[width=0.49\linewidth]{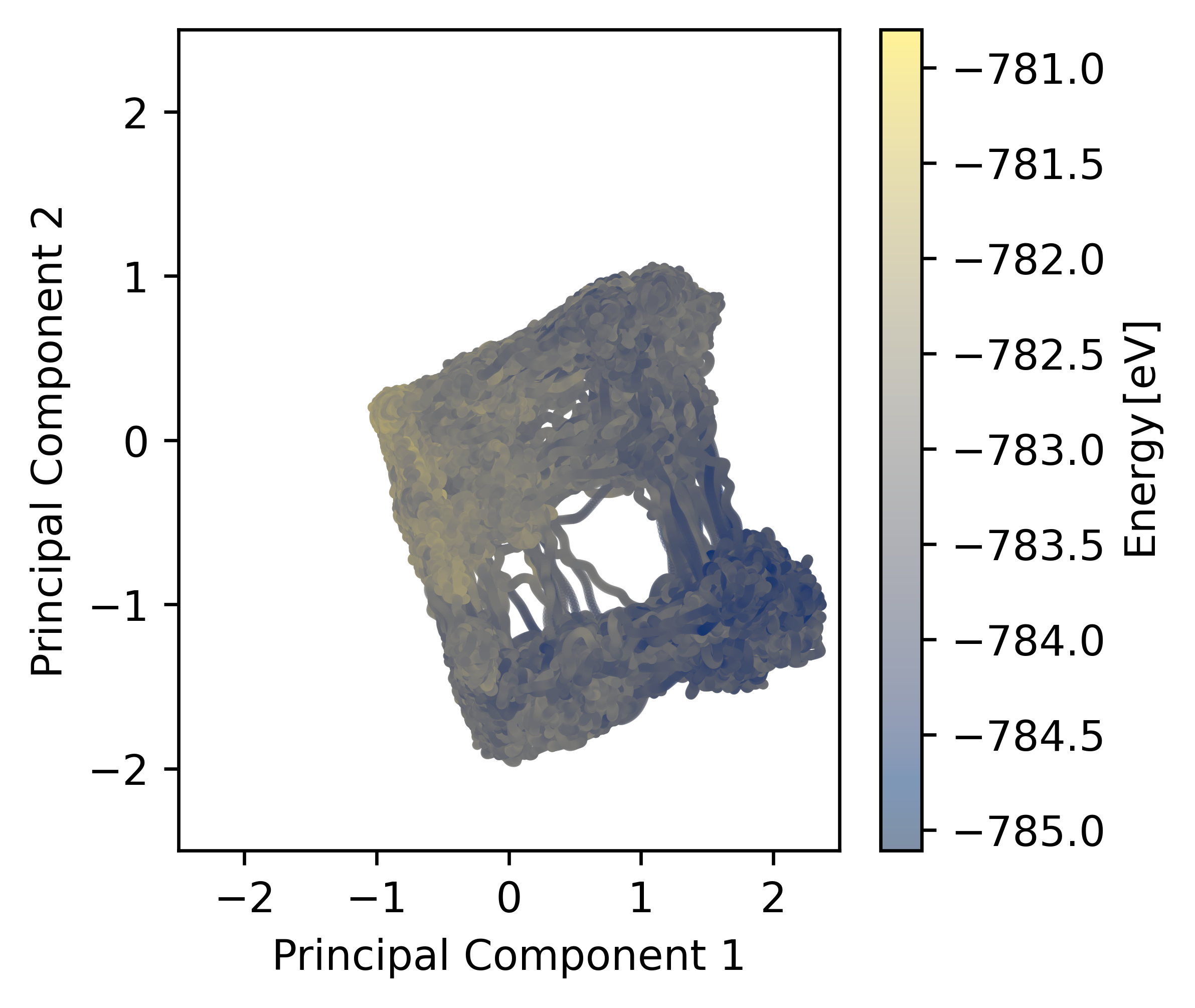}
    \caption{PCA analysis of 100 (left) and 1000 (right) ML trajectories (20.000 steps per trajectory).}
    \label{Sfig:PCA_1000}
\end{figure}


\section{Unfunctionalized Cyclooctyne}
\begin{figure}[h]
    \centering
    \includegraphics[width=0.9\columnwidth]{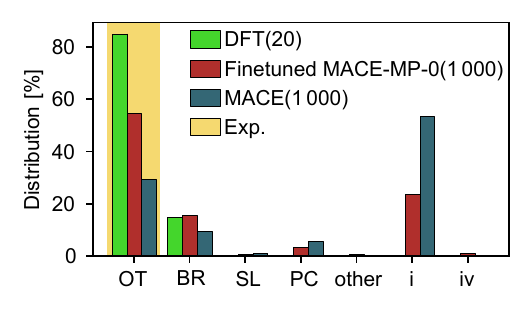}
    \caption{Distribution of adsorption sites of unfunctionalized cyclooctyne on Si(001) using the model fine-tuned on ECCO (FT) as well as the model trained from scratch (FS).}
    \label{Sfig:barplot_COC_Si}
\end{figure}
To estimate the transferability of the fine-tuned and from-scratch models, we perform MD simulations using both models on a slightly different system: the \textbf{un}functionalized cyclooctyne molecule on the Si(001) surface. For this system, there is again previous available AIMD data from the group of RTZ.\cite{Pecher2017c} The MD setup is the same as for the ECCO molecules, see the main text for details.

The fine-tuned model performs well, none of our MD simulations show any signs of breaking of the cyclooctyne molecules, that is often encountered during ML driven MDs (see also Figure \ref{Sfig:broken molecules}). The resulting adsorption behaviour reproduces the trends from the AIMD study of Pecher \textit{et al.}\cite{Pecher2017c} well (see Figure \ref{Sfig:sampling_COC_on_Si}. However, more bridge and sub-layer adsorption modes are found than occur in the DFT reference. As shown in the main text, this might however only be due to the limited amount of AIMD trajectories.

In comparison, the from-scratch model performs, as expected, significantly worse. Mainly, this is due to an increase in desorbed molecules. The reduced amount of breaking ECCO molecules could be explained by the longer training time, that enables the from-scratch model to better represent strained variants of the ECCO molecule.
\begin{figure}[h]
     \centering
    \includegraphics[width=0.9\columnwidth]{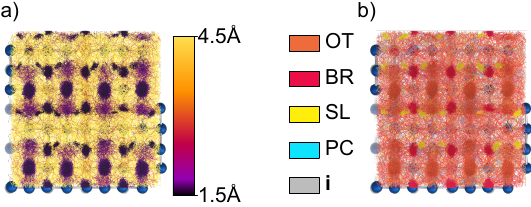}
    \caption{Binning of the measured lowest triple bond position at each xy-bin. a) color is representing the height of the measured points and b) the adsorption site of the respective trajectories.}
    \label{Sfig:sampling_COC_on_Si}
\end{figure}

\providecommand*{\mcitethebibliography}{\thebibliography}
\csname @ifundefined\endcsname{endmcitethebibliography}
{\let\endmcitethebibliography\endthebibliography}{}